\documentclass{aa}
\usepackage{txfonts}
\usepackage{amsmath}
\usepackage{color}
\usepackage{xcolor}
\usepackage[normalem]{ulem}
\usepackage{multicol}
\usepackage[colorlinks=true,citecolor=blue,linkcolor=blue]{hyperref}
\usepackage{textcomp}
\usepackage{subcaption}
\usepackage{graphicx}
\usepackage{float}
\usepackage{placeins}
\usepackage{multirow}
\usepackage{changepage}
\usepackage{colortbl}
\usepackage{amsfonts}
\usepackage{booktabs}
\usepackage{siunitx}
\usepackage{amssymb}
\usepackage{enumitem}
\usepackage{multirow}
\usepackage{gensymb}
\usepackage{textcomp}
\usepackage{upgreek}
\usepackage{stfloats}
\usepackage{comment}

\author{
J.M.G.H.J. de Jong\inst{\ref{inst:Leiden},\ref{inst:Astron}} 
\and L.K. Morabito\inst{\ref{inst:Durham},\ref{inst:Durham2}}
\and R.J. van Weeren\inst{\ref{inst:Leiden}}
\and J.B.R. Oonk\inst{\ref{inst:surf},\ref{inst:Leiden}}
\and A. Ahmadi\inst{\ref{inst:Astron}}
\and P.N. Best\inst{\ref{inst:edin}}
\and C.L. Hale\inst{\ref{inst:edin}}
\and R. Kondapally\inst{\ref{inst:Durham},\ref{inst:Durham2}}
\and G. Lusetti\inst{\ref{inst:open}}
\and H.J.A. R\"ottgering\inst{\ref{inst:Leiden}}
\and T.W. Shimwell\inst{\ref{inst:Astron},\ref{inst:Leiden}}
\and F.~Sweijen\inst{\ref{inst:Durham}}
\and R. Timmerman\inst{\ref{inst:Durham},\ref{inst:Durham2}}
\and M. van der Wild\inst{\ref{inst:Durham}}
}

\institute{
Leiden Observatory, Leiden University, PO Box 9513, 2300 RA Leiden, The Netherlands\relax\label{inst:Leiden} 
\and 
ASTRON, The Netherlands Institute for Radio Astronomy, Postbus 2, 7990 AA Dwingeloo, The Netherlands\relax\label{inst:Astron}
\and
Centre for Extragalactic Astronomy, Department of Physics, Durham University, Durham, DH1 3LE, UK\relax\label{inst:Durham} 
\and 
Institute for Computational Cosmology, Department of Physics, Durham University, South Road, Durham DH1 3LE, UK\relax\label{inst:Durham2}
\and 
SURF/SURFsara, Science Park 140, 1098 XG Amsterdam, The Netherlands\relax\label{inst:surf} 
\and
Institute for Astronomy, University of Edinburgh, Royal Observatory, Blackford Hill, Edinburgh, EH9 3HJ, UK \relax\label{inst:edin}
\and
School of Physical Sciences, The Open University, Walton Hall, Milton Keynes, MK7 6AA, UK\relax\label{inst:open} 
}

\begin{document}

\title{Pushing the frontiers of low-frequency radio astronomy: Ultra-deep sub-arcsecond LOFAR images of ELAIS-N1}
\titlerunning{Pushing the frontiers of low-frequency radio astronomy}

\begin{abstract} 
    {We present ultra-deep sub-arcsecond radio images of the ELAIS-N1 field covering $2.5\times2.5$~deg$^{2}$ produced from 200~hours of LOw Frequency ARray (LOFAR) observations in the 115–168 MHz band. The radio continuum images were obtained using recent advances in LOFAR long-baseline calibration and imaging. We achieved a central point-source sensitivity of 5.3~$\upmu$Jy~beam$^{-1}$ at an angular resolution of 0.33\arcsec$\times0.39$\arcsec\ and an area-weighted sensitivity of 7.4~$\upmu$Jy~beam$^{-1}$ at 0.36\arcsec$\times0.45$\arcsec, making these the deepest radio images below 1.28~GHz to date and among the deepest at any radio frequency. By combining our data with previously released deep 6\arcsec~LOFAR images, we have compiled a catalogue with a total of 
    62\,916 unique radio sources across $2.5\times2.5$~deg$^{2}$. Our calibration strategy, combined with astrometric alignment using 95 \textit{Gaia} quasars, resulted in positional offsets of $\Delta$RA = $-3 \pm 27$~mas\ and $\Delta$DEC = $2 \pm 17$~mas for these same sources. Further, our work demonstrates the potential for deep polarisation studies at sub-arcsecond resolution with instrumental leakages from Stokes I into linear polarisation down to 0.12\%, which is about a factor of two better than the typical leakage in the LOFAR Two-metre Sky Survey at 6\arcsec\ resolution. Our results demonstrate that imaging a large number of observations over a wide field area at sub-arcsecond resolution is now computationally feasible. This establishes a new benchmark for low-frequency radio imaging in terms of combining depth, resolution, and survey area, which lays the foundation for deeper and wider surveys in the LOFAR2.0 and SKA era.}
    \maketitle
\end{abstract}

\section{Introduction}

The LOw Frequency ARray \citep[LOFAR;][]{haarlem2013} is currently the only interferometer capable of routinely producing sub-arcsecond imaging of the low-frequency radio sky \citep[]{morabito2022}. Using LOFAR's high-band antennas (HBAs), we have over the past few years carried out wide-field (2.5$\times$2.5 deg$^{2}$) imaging of several target fields at sub-arcsecond angular resolution in the 115--168 MHz band by utilising baselines up to 1\,980~km across the European continent \citep{sweijen2022, dejong2024a, escott2026}. This unique capability has enabled studies of a broad range of science cases that require sub-arcsecond resolution over wide areas of the sky \citep[for a review, see][]{morabito2025a}. Extending these capabilities to produce ultra-deep sub-arcsecond images of the LOFAR deep fields \citep[see][for current deep fields]{sabater2021, tasse2021, best2023, bondi2024, shimwell2025} is a next step in order to enabling the detection and study of the faintest and most distant radio source populations.

The sub-arcsecond wide-field imaging capability of LOFAR was for the first time shown to be possible by \citet{sweijen2022}, who produced the first wide-field image with one 8-hour observation. This was followed by a demonstration of deeper imaging by \cite{dejong2024a} using 32 hours of observations. While these studies highlighted the potential of sub-arcsecond wide-field imaging with LOFAR, scaling up to ultra-deep imaging with hundreds of hours of LOFAR observations remained non-trivial. The primary challenges included processing and imaging 16~TB of visibility data per 8-hour observation\footnote{These data are typically reduced to around 4~TB using dynamical statistical compression \citep[Dysco;][]{offringa2016}, although the total number of visibilities remains unchanged.} within a computationally feasible timescale, automatically selecting and precisely calibrating high signal-to-noise ratio (S/N) calibrator sources across the wide field of view (FoV) under varying ionospheric conditions, the difficulty of jointly calibrating the short baselines from the Dutch stations with the long baselines from the European stations, and the need for human intervention for validation and quality assessment.

Recent work has addressed the aforementioned challenges. For instance, there have been general developments in the calibration software \texttt{facetselfcal} \citep{vanweeren2021} that have improved calibration performance and reliability \citep[e.g.][Appendix~A]{vanweeren2026a} when also jointly calibrating short and long baselines \citep{dejong2025b}. Automated self-calibration parameter selection and quality assessment based on a combination of quality metrics with a machine learning algorithm have removed the need for human intervention \citep{dejong2024a, dejong2025b}. The high computational costs for the final deep imaging with multiple observations are mitigated through sidereal visibility averaging (SVA; \citealt{dejong2025}). This approach averages calibrated visibilities from different epochs with similar baseline coordinates, which, after applying all calibration solutions, reduces the total data volume of tens of observations to that of one or two. Combining this with recent improvements in automated processing strategies \citep{dejong2025b}, the overall computational cost of processing for ultra-deep imaging with hundreds of hours of LOFAR data has been lowered by more than an order of magnitude compared to the first successful LOFAR sub-arcsecond wide-field imaging by \cite{sweijen2022}. The use of the Common Workflow Language \citep[CWL;][]{amstutz2016, crusoe2022}, Toil \citep{vivian2017}, and Simple Linux Utility for Resource Management \citep[SLURM;][]{andy2003} for data processing on high-performance computing clusters, have also enabled efficient automated parallel processing of numerous observations, as detailed in \cite{pilot}. The CWL workflows are reusable, interoperable, and portable, which aligns CWL-based pipelines with FAIR principles \citep[e.g.][]{wilkinson2016} and supports scalable processing of large numbers of LOFAR observations.

In this work, we use the most recent data processing strategies for LOFAR long-baseline imaging to release ultra-deep sub-arcsecond wide-field images across a FoV of $2.5\times2.5$~deg$^{2}$ of the European Large Area ISO Survey-North 1 \citep[ELAIS-N1;][]{oliver2000} deep field using 200~hours of LOFAR-HBA observations in the 115--168~MHz band. 
The highest resolution image represents the deepest LOFAR image to date and is among the deepest radio images ever produced at any frequency.
Located at a right ascension (RA) of 16:11:00 and a declination (DEC) of +54:57:00, ELAIS-N1 has been extensively studied across a broad range of wavelengths, including X-rays \citep[e.g.][]{manners2003}, ultraviolet \citep[e.g.][]{martin2005, chaturvedi2026}, optical \citep[e.g.][]{mcmahon2001, aihara2018}, infrared \citep[e.g.][]{lawrence2007, mauduit2012, oliver2012}, and radio \citep[e.g.][]{sirothia2009, ocran2020, sinha2022, kondapally2022, best2023}, and it is one of the selected LOFAR deep fields \citep{kondapally2021, duncan2021, tasse2021, sabater2021, bondi2024}. The deepest sub-arcsecond LOFAR image so far of this field was produced by \citet{dejong2024a} using 32~hours of LOFAR observations, reaching a central sensitivity of 14~$\upmu$Jy~beam$^{-1}$. More recently, \citet{shimwell2025} imaged the same field using 505 hours of all available LOFAR observations containing data from the Dutch stations, achieving a central sensitivity of about 10~$\upmu$Jy~beam$^{-1}$ at 6\arcsec\ resolution. Although they used about 16 times more observing time than \cite{dejong2024a}, the two images reach comparable point-source sensitivities. This is a consequence of both the larger collecting area of the international stations \citep{haarlem2013} and of the confusion noise \citep{condon2012}, which limits deep 6\arcsec-imaging with LOFAR with only the baselines from the Dutch stations \cite[e.g.][]{sabater2021, shimwell2025}.

Combined with the wealth of existing multi-wavelength data \citep{kondapally2021, duncan2021} and the forthcoming deep optical–infrared catalogues of ELAIS-N1, our ultra-deep sub-arcsecond images open a new observational window onto a broad range of science cases. These include separating star formation and active galactic nuclei (AGN) contributions to the radio emission of fainter sources \citep{morabito2022a, morabito2025}, probing the cosmic star formation history to high redshifts \citep[e.g.][]{novak2017, cochrane2023, wang2024,thykkathu2026}, extending measurements of the radio luminosity function with a more complete sample of the faintest and least powerful radio jets \citep[e.g.][]{gendre2010, novak2018, kondapally2022, dejong2024b}, among other applications \citep[e.g.][]{morabito2025a}. Our results also lay the foundation for even deeper and wider future surveys that aim to exploit the full international LOFAR during the upcoming LOFAR~2.0 era.

This paper is structured as follows. In Sect.~\ref{sec:datareduction} we describe the selection of the ELAIS-N1 observations used in this work and the data processing strategy to reduce the data. In Sect.~\ref{sec:cataloguing}, we outline the cataloguing process for obtaining a single multi-resolution source catalogue. The data quality is presented in Section~\ref{sec:results}, followed by a discussion and the conclusions in Sects.~\ref{sec:discussion} and~\ref{sec:conclusions}, respectively.

\section{Observations and data reduction}\label{sec:datareduction}

The data reduced in this work cover the ELAIS-N1 deep field (RA~$=$~242.75~deg, DEC~$=$~54.95~deg). Below we describe the data selection and processing workflow used to reduce the data. We also discuss the flux density scale and astrometric corrections applied to the data prior to imaging and analysis.

\subsection{Data selection}\label{sec:data}

Out of the total 505~hours of available LOFAR observations, targeting the ELAIS-N1 field \citep{shimwell2025}, we selected 200~hours from the \texttt{LT10\_012} and \texttt{LT14\_003} projects (PI: P.N. Best). The corresponding 25 LOFAR observations are listed in Appendix \ref{app:observations}. Hence, 305~hours were excluded based on the following criteria:
\begin{itemize}
    \item All observations from projects \texttt{LC2\_024} and \texttt{LC4\_008} were recorded either with Dutch LOFAR stations only or with international stations excluding the most distant ones (in Poland, Latvia, and Ireland), which were not yet part of the array when these data were taken \citep{polfar2016,latvia2020,murphy2021}. This makes them unsuitable for sub-arcsecond resolution imaging.
    \item Four observations were recorded at 10~sec intervals. This time resolution results in extreme time smearing effects when imaging at sub-arcsecond resolution (see Section \ref{sec:smearing}).
    \item Three observations showed elevated RMS noise levels, exceeding twice the expected values at 6\arcsec~resolution as reported by \citet{shimwell2025}, making them unsuitable for calibration with the international stations.
    \item Eight observations were excluded after visual inspection of the calibration solutions from the LOFAR Initial Calibration pipeline \citep[\texttt{LINC};][]{degasperin2019}, due to excessive flagging (>40\%) of the international stations.
\end{itemize}  
The observations from project \texttt{LT10\_012} were stored with a 1-second time resolution, whereas those from \texttt{LT14\_003} used a 2-second resolution, which increases time-smearing at the edge of the FoV (see Section \ref{sec:smearing}). Additionally, the \texttt{LT14\_003} pointings were offset by 0.03\degree~from the field centre from \texttt{LT10\_012}, requiring a phase shift to align all observations to a common RA of 16:11:00 and DEC of +54:57:00. The observations \texttt{L686962}, \texttt{L769393}, \texttt{L798074}, and \texttt{L816272} had already been processed by \citet{dejong2024a}, but were reprocessed in this work at the DD calibration step using the updated calibration strategy by \citet{dejong2025b}.

\subsection{Calibration}\label{sec:calibration}

\begin{figure}[htbp]
 \centering
\includegraphics[width=1\linewidth]{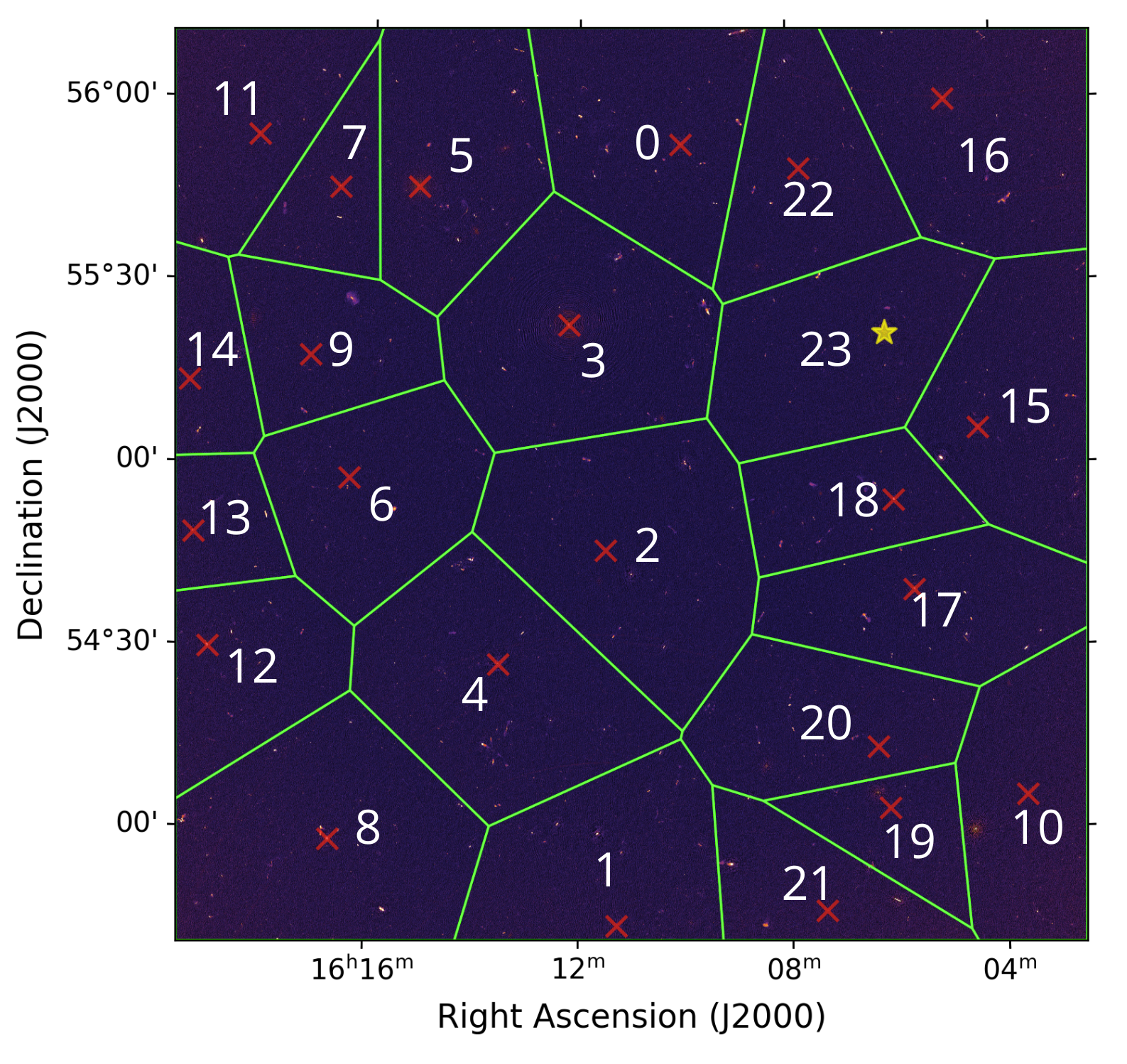}
  \caption{Facet layout of the 24 calibrators with the corresponding facet IDs. The DI calibrator is indicated by the yellow star, and the other DD calibrators are marked with red crosses. The facet boundaries are indicated by the green lines and are based on a Voronoi tessellation, which divides the field into polygon regions centred on calibrator sources, where each point in the field is assigned to its nearest calibrator. The calibration solutions are assumed to be constant within each polygon region \citep{schwab1984, vanweeren2016}.}
\label{fig:facet_layout}
\end{figure}

To calibrate the selected datasets, we adopted the strategy outlined in \cite{dejong2024a} up to the direction-independent (DI) calibration stage. Beyond this point, we followed the direction-dependent (DD) calibration strategy of \cite{dejong2025b} for Stokes I and the polarisation alignment strategy of \cite{vanweeren2026} for deep polarisation imaging. This means that after downloading the data from the LOFAR long-term archive \citep[LTA;][]{renting2012,haarlem2013}, we applied the HBA calibrator and target workflows from \texttt{LINC}\footnote{\url{https://linc.readthedocs.io}}, the \texttt{DDF}-pipeline\footnote{\url{https://github.com/mhardcastle/ddf-pipeline}}, workflows from the Pipeline for the International LOFAR Telescope (\texttt{PILOT})\footnote{\url{https://github.com/LOFAR-VLBI/pilot}}, and additional calibration steps with \texttt{facetselfcal}\footnote{\url{https://github.com/rvweeren/lofar_facet_selfcal}} in the order outlined below. As each processing step has been described in detail in previous work, we refer the reader to the corresponding cited papers for further details. In summary, the overall processing steps were applied as follows:
\begin{enumerate}
\item The first step involved executing the automated \texttt{LINC} workflows, which derived initial calibration corrections for instrumental and ionospheric effects across all LOFAR stations \citep{vanweeren2016, degasperin2019}. These corrections were obtained from a 10-minute scan of the well-known calibrator source 3C~295, while additional calibration solutions for the Dutch LOFAR stations were derived from the ELAIS-N1 target field data.
\item Next, we ran the \texttt{DDF}-pipeline for DD calibration of the Dutch LOFAR stations \citep{tasse2014, tasse2014b, tasse2018, tasse2021, smirnov2015}. This gave us full-Jones DI calibration solutions for the Dutch LOFAR stations, which we used as a starting point for the calibration of the international LOFAR stations, and DD calibration solutions to subtract sources located outside the 2.5$\times$2.5~deg$^{2}$ FoV \citep{sweijen2022}, after which the DD solutions were discarded.
\item After obtaining the calibration corrections for the Dutch stations, we used the \texttt{delay-calibration} workflow from \texttt{PILOT} to obtain full-Jones DI calibration corrections for all LOFAR stations \citep{morabito2022}. We used the Seyfert 2 galaxy ICRF\,J160607.6+552135, selected by \cite{ye2024} from the candidate calibrators from the Long-Baseline Calibrator Survey \citep[LBCS,][]{jackson2016, jackson2022}, as calibrator source. This calibration step incorporated the skymodel and refined calibration steps described in \cite{dejong2024a}. Because the source is unpolarised \citep{ruiz2021, piras2024}, any polarised signal detected towards it is instrumental, so with the full-Jones calibration the polarisation leakage was corrected.
\item After applying the DI corrections to our data, we performed an additional DD calibration step with \texttt{facetselfcal} to derive solutions for the Dutch LOFAR stations in several high S/N directions. This is essential for stabilising the subsequent DD calibration of the Dutch and international stations combined. We followed the strategy of \cite{dejong2025b}, which describes this step in more detail, with one main difference: we first jointly calibrated and imaged six observations (the 4 observations from \cite{dejong2024a} with \texttt{L686892} and \texttt{L686906}) to produce an accurate sky model of the full 2.5$\times$2.5~deg$^2$ FoV, against which we then calibrated all remaining observations in a single iteration. This step is not yet part of \texttt{PILOT}, but may be added in the future.
\item Next, with the datasets with the DI calibration solutions for the international stations applied and using the DD calibration solutions for the Dutch stations, we applied the \texttt{dd-calibration} workflow from \texttt{PILOT} to obtain DD calibration solutions for all LOFAR stations, as detailed in \cite{dejong2025b}. This workflow pre-applied the nearest calibration solutions from step 4 and then calibrated the international LOFAR stations. See Figure \ref{fig:facet_layout} for the DD calibrator selection and corresponding facet layout, where the calibration solutions are assumed to be constant within each polygon from the Voronoi tessellation \citep{schwab1984, vanweeren2016}.
\item With the calibration solutions for all international stations we made the first high-resolution wide-field images with the \texttt{image-intermediate-resolution} workflow from \texttt{PILOT}. This returned images of the full FoV at an angular resolution of 1.2\arcsec~\citep[e.g.][]{ye2024,dejong2024a}. Most important data products from this step were the fully calibrated Stokes I images for quality assessment (see RMS noise levels in Appendix \ref{app:observations}) and the model images from each frequency channel, as these were used as input for the subsequent processing step \citep{dejong2024a}. 
\item Using the 1.2\arcsec~model images, we ran the \texttt{facet-subtract} workflow from \texttt{PILOT} to split each facet into its own visibility dataset, where all sources outside of the facet were subtracted \citep{dejong2025b}.
\item After the facet-subtraction, we applied a final calibration step, in which the selected calibrator sources from step 5 were calibrated jointly across all observations. This followed the method described in \cite{dejong2025b}. This ensured all observations were astrometrically aligned before the final combined imaging and further improved the deep calibration quality as signal from other bright sources in the field was here subtracted outside the facets. The calibration solutions were directly applied to the facet data. The workflow to perform this will be added in the future to \texttt{PILOT}.
\item Finally, to allow for deep polarisation imaging, we aligned the polarisation angles between the different observations, using the known polarised source 7C~1604+5447 \citep{ruiz2021, piras2024}. The steps for this alignment procedure and for this particular source are detailed in \cite{vanweeren2026}.
\end{enumerate}
The advantage of this calibration strategy is that all 25 observations are processed independently up to step 8 (with the exception of the additional step 4). The joint calibration in step 8 is performed only after the preceding processing steps have been completed for all observations. This enables embarrassingly parallel processing of the computationally intensive calibration and imaging steps up to that point.\footnote{A workload is termed `Embarrassingly parallel' when its subtasks are mutually independent and thus require no inter-process communication.} The RMS noise levels of the 1.2\arcsec~wide-field images produced in step 6 range from 81 to 126 $\upmu$Jy beam$^{-1}$, as shown in Appendix \ref{app:observations}. This reflects the combined effects of data flagging, ionospheric conditions, and variations in instrumental/observing properties across the years the observations were taken. After all calibration steps, we had datasets with fully DD calibrated visibilities for each of the 24 facets. These were then imaged, as outlined in the next subsection.

All the (self-)calibration steps, with \texttt{PILOT} and during the extra calibration steps, were done with \texttt{facetselfcal}. This software package utilises the Default Preprocessing Pipeline (\texttt{DP3}; \citet{dp3, dijkema2023}) and \texttt{WSClean} \citep{wsclean, offringa2017}, and functionalities from \texttt{lofar\_helpers} \citep{lofarhelpers},\footnote{\url{https://github.com/jurjen93/lofar_helpers}} to perform (self-)calibration on a source. Recent enhancements to \texttt{facetselfcal} have improved the reproducibility of calibration through improved metadata logging and features to return configuration files from calibration solutions and images. These developments represent an important step towards aligning the software with FAIR principles \citep[e.g.][]{wilkinson2016, dejong_fair}. This is essential to keep track of automatically selected parameters and to reproduce results in case further improvement of the output data or deeper imaging is desired in the future (see Section \ref{sec:future}). All software was executed with Apptainer containers \citep{kurtzer2017} from \texttt{FLoCs}\footnote{\url{https://tikk3r.github.io/flocs}} \citep{flocs600}.

\subsection{Imaging}\label{sec:imaging}

\begin{figure*}[htbp]
 \centering
\includegraphics[width=0.88\linewidth]{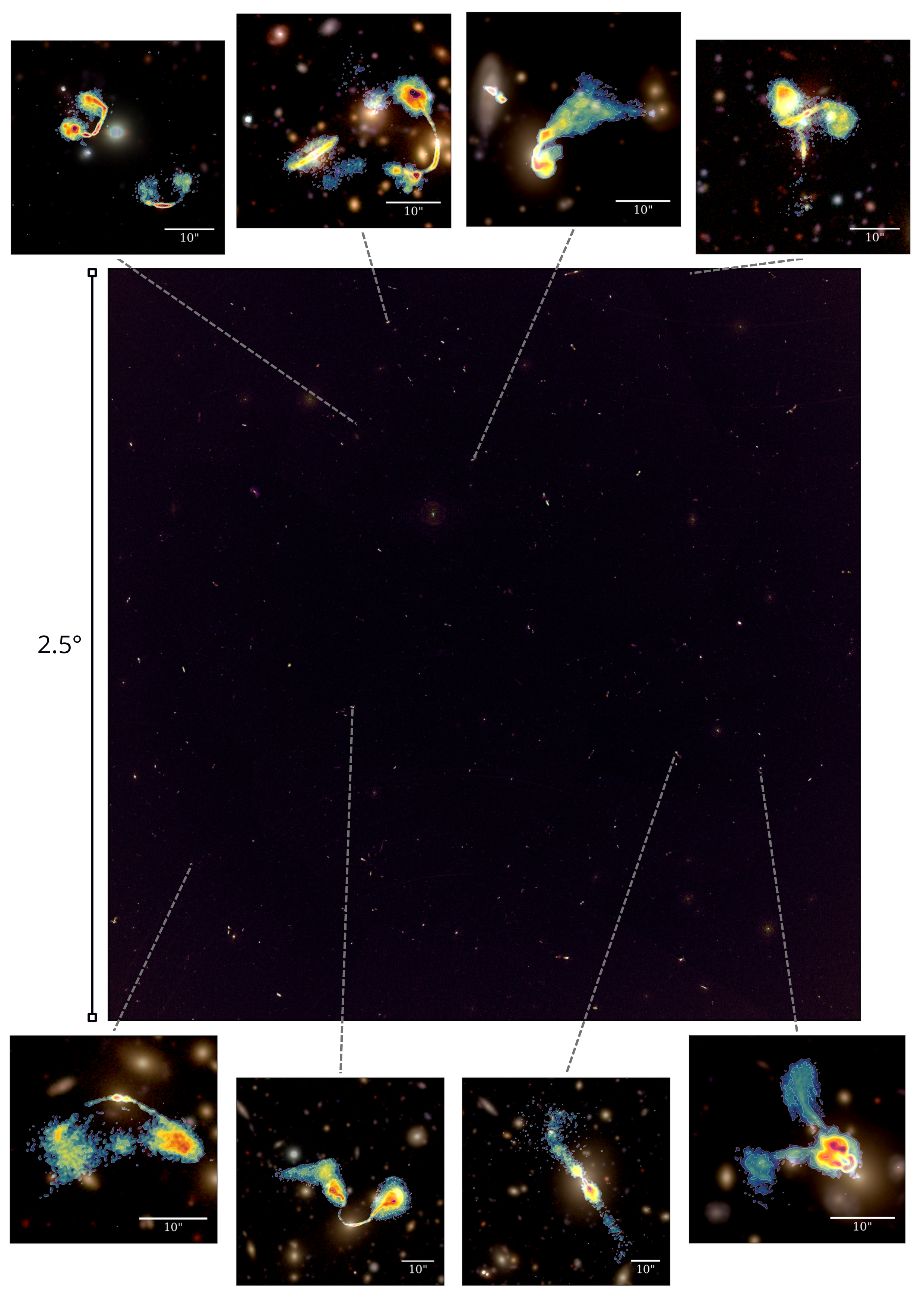}
  \caption{Full wide-field radio continuum LOFAR image of the ELAIS-N1 field at 1.0\arcsec$\times1.5$\arcsec~angular resolution with postage stamp images of various radio galaxies at 0.36\arcsec$\times0.45$\arcsec~resolution. The optical background is a three-colour composite of the Subaru Hyper Suprime-Cam \citep[HSC;][]{aihara2018}, where $g$-, $r$-, and $i$-band imaging are mapped to blue, green, and red, respectively. The colour map for radio emission was generated with \texttt{CMasher} \citep{cmasher}. We show several resolution comparisons of these sources in Appendix~\ref{appendix:6vs0.3}.}
\label{fig:layout}
\end{figure*}

To reduce the computational cost for deep imaging combining the datasets from each individual observation, we applied SVA after calibration \citep{dejong2025}. This method averages the visibilities from different epochs with similar baseline coordinates. Since the baseline coordinates (\textit{uvw}) in the datasets are set by the time axis, the accuracy of SVA is determined by the time resolution, meaning that only the time resolution for the output dataset needs to be specified. Before SVA, each facet dataset was averaged in time by a factor set by the size of the facet \citep{dejong2025b}. To minimise the introduction of additional time-smearing effects, we over-sampled the time axis of the final SVA-applied dataset by a factor of two compared to the averaged facet datasets, based on the analysis from \cite{dejong2025}. In this way, we reduced the total number of calibrated and facet-subtracted visibilities from all observations combined (about 200~TB of Dysco-compressed visibilities) by a factor of 12.5, which decreased the computational cost of gridding and prediction during imaging by about an order of magnitude compared to imaging without SVA.

Imaging was performed with \texttt{WSClean} \citep{wsclean}, following the imaging parameters from \cite{dejong2024a} and using the \texttt{wgridder} module \citep{arras2021, ye2022}. Hence, for each facet, we used a minimum \textit{uv}-cut of 80$\lambda$, and pixel sizes of 0.1\arcsec, 0.2\arcsec, and 0.4\arcsec. These were paired with Gaussian tapers of 0.3\arcsec, 0.6\arcsec, and 1.2\arcsec, respectively. The Briggs weighting \citep{briggs1995} was set by \cite{dejong2024a} to $-1.5$. However, for the highest angular resolution we changed this value in this work to $-1.35$ as this improved the point-source sensitivity by about 10\%, while still achieving high angular resolution.
To ensure efficient deconvolution, we employed automatic masking, multi-scale cleaning \citep{cornwell2008}, and an RMS box size set to 50 times the synthesised beam \citep{offringa2017}. We also applied a correction for the primary beam attenuation. 

Each individually imaged facet has its own restoring beam and RMS noise, as reported in Appendix~\ref{app:facetproperties}. The labels 0.3\arcsec, 0.6\arcsec, and 1.2\arcsec\ refer to the Gaussian tapers used during imaging, while the actual restoring beams of each facet, as fitted by \texttt{WSClean}, are listed in this table. Facets further from the pointing centre have larger restoring beams at all three resolutions. They reflect the combination of different LOFAR station beams. The international stations are larger than the Dutch stations and therefore have narrower station beams, so away from the phase centre their sensitivity drops faster than that of the Dutch stations. Because the beam correction applied to each facet dataset also rescales the visibility weights, the longest baselines receive relatively lower weights in these facets, which broadens the PSF. The same primary beam attenuation from all stations combined also causes the RMS noise to increase towards the edge of the field. Before mosaicing the facets together, we restored them to a common resolution of $0.36\arcsec\times0.45\arcsec$, $0.58\arcsec\times0.62\arcsec$, and $1.0\arcsec\times1.5\arcsec$, matching the resolutions used by \cite{dejong2024a}. The mosaicing was done with \texttt{SWarp}\footnote{\url{https://www.astromatic.net/software/swarp}} \citep{swarp}. For many facets, in particular the ones labelled with 1.2\arcsec, the restoring beam is smaller than the native beam. Because the residuals retain the native PSF, the flux density units are then not consistent within a facet. The mosaics are therefore intended for visualisation purposes only. All catalogues in this work were extracted from the individual facet images at their native resolution (see Section~\ref{sec:cataloguing}), and quantitative measurements should be made on these facet images, which we release alongside the mosaics.

In Figure \ref{fig:layout} we display the full mosaic at 1.0\arcsec$\times1.5$\arcsec\ angular resolution with postage stamps of radio galaxies at 0.36\arcsec$\times0.45$\arcsec\ angular resolution. In Appendix \ref{appendix:6vs0.3} we demonstrate with a few postage stamp images the resolving power of imaging at different angular resolutions.

\subsection{Flux scale alignment}\label{sec:fluxdensityscaling}

Because of our extensive DD calibration process (steps 4, 5, and 8 in Section~\ref{sec:calibration}), there is an increased likelihood that the amplitude corrections introduced small drifts in the flux density scale that was accurately set in step 3 from Section~\ref{sec:calibration}. For this reason, and to align our flux density scale with the recent 6\arcsec\ results of \citet{shimwell2025}, we imaged the fully calibrated facet datasets at 6\arcsec\ resolution, including the international stations and using the imaging parameters from Section~\ref{sec:imaging} with a 6\arcsec\ Gaussian taper. We then generated a source catalogue using the source finder \texttt{PyBDSF} \citep[][see Section~\ref{sec:cataloguing} for details]{mohan2015}. We cross-matched our 6\arcsec\ catalogues for each facet with theirs for sources having peak intensities larger than ten times the local RMS and with integrated flux density measurements larger than five times the reported \texttt{PyBDSF} error. This provided us with 5\,379 sources which we used to align the flux scales.

Dividing the flux densities from \citet{shimwell2025} by our measurements at the same resolution yielded a median flux density ratio of $0.95\pm0.02$, with values ranging from $0.89\pm0.01$ to $1.07\pm0.01$ across the individual facets. The uncertainties are the standard error of the median, obtained by bootstrap resampling with each flux density perturbed within the uncertainty reported by \texttt{PyBDSF}. \citet{shimwell2025} used different software and imaging parameters (\texttt{DDFacet}; \citealt{tasse2018}, Dutch stations only, Briggs robust $-0.5$, and no taper). Since the integrated flux densities of high S/N compact sources are, to first order, insensitive to the visibility weighting scheme and deconvolution depth, we expect these differences to have only a minor effect. They may contribute to a small global offset, but they cannot explain the facet-to-facet variation, as our imaging parameters are identical for all facets. For most facets, the ratios are also within the $\sim$9\% flux density scale uncertainty of \citet{shimwell2025}.

Applying the facet-to-facet scale factors using the median ratios brought our flux density scale into closer agreement with that of \citet{shimwell2025}, yielding an average flux density ratio of $1.00\pm0.01$ across the full FoV. This procedure assumes the flux density scale of \citet{shimwell2025} to be free of facet-to-facet variations. Residual DD errors at the percent level cannot be excluded in their data either, so our correction should be understood as placing our images on the same scale as \citet{shimwell2025} rather than as an absolute flux density calibration. To demonstrate the accuracy of our flux density scale, we show in Appendix~\ref{appendix:spectra} the spectra of several of our DD calibrators across measurements from different radio surveys.

\subsection{Astrometric corrections}\label{sec:astrometry}

The previous work by \cite{dejong2024a} reported a median astrometric offset of $\Delta$RA = $94 \pm 93$~mas and $\Delta$DEC = $-67 \pm 64$~mas, based on comparing their radio sources positions with optical positions from \cite{kondapally2021}, which were calibrated to the Two Micron All Sky Survey \citep[2MASS;][]{skrutskie2006}. Since the astrometric accuracy of 2MASS is around 100~mas, these data were of insufficient quality to further improve the astrometric precision. To obtain a better estimate of the astrometric precision, in this work we instead used the positions of quasars detected by \textit{Gaia} \citep{gaiadr3}, which have sub-milliarcsecond positional accuracy.

\begin{figure}[htbp]
 \centering
\includegraphics[width=0.49\linewidth]{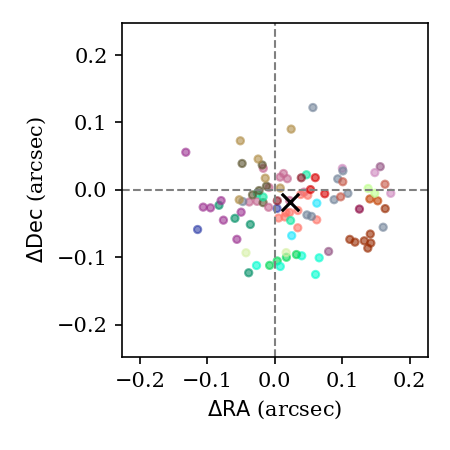}
\includegraphics[width=0.49\linewidth]{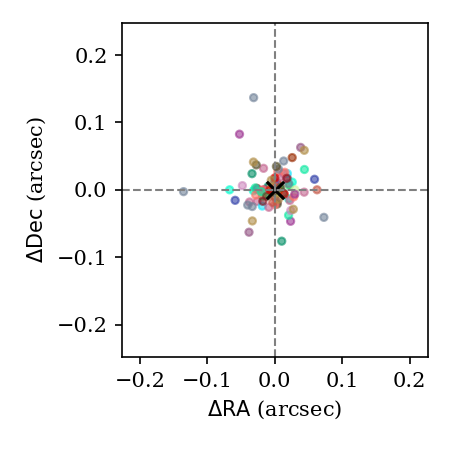}
  \caption{Astrometric accuracy of 95 \textit{Gaia} quasars relative to the nearest radio sources in our newly obtained dataset before (\textit{left}) and after (\textit{right}) astrometric corrections. The black cross marks the median offsets in $\Delta$RA and $\Delta$DEC, while different facets are shown in distinct colours to highlight the facet-to-facet offsets. For sign convention, we used RA$_{\text{Gaia}}$-RA$_{\text{LOFAR}}$ and DEC$_{\text{Gaia}}$-DEC$_{\text{LOFAR}}$.}
\label{fig:astrometry}
\end{figure}

We used \texttt{PyBDSF} to determine the positions of the sources in our highest resolution facet images (see Section \ref{sec:cataloguing}). To select quasars from our radio detections, we used the \texttt{optAGN} column from \cite{kondapally2021}, which contains sources from the Million Quasar Catalog \citep{flesch2023} or which were spectroscopically identified as AGN. These sources were cross-matched with our radio positions within 6\arcsec, which matched with the radio catalogue resolution from the cross-matched catalogue. This left us with 260 sources that we then matched with the nearest \textit{Gaia} source within 1\arcsec. This resulted in 95 source matches across 22 of our 24 facets, where we found a global median astrometric offset of $\Delta$RA = $22 \pm 71$~mas\ and $\Delta$DEC = $-17 \pm 41$~mas. The uncertainty is the scaled median absolute deviation (SMAD).

In the left panel of Figure \ref{fig:astrometry}, we show the $\Delta$RA and $\Delta$DEC for these sources, with different facets indicated by different colours. The offsets cluster together for many of the facets, which can be explained by the fact that each facet received its own phase corrections during DD calibration, resulting in facet-specific systematic astrometric offsets. To correct for this, we applied facet-to-facet astrometric corrections based on the offsets relative to the \textit{Gaia} positions. For facets 8 and 14, there were no \textit{Gaia} quasar matches, so we applied to these facets the global median astrometric corrections from all facets together. After applying all the corrections, the median astrometric offsets for the 95 quasars are reduced to $\Delta$RA $= -3 \pm 27$~mas and $\Delta$DEC $= 2 \pm 17$~mas (excluding facets 8 and 14), as shown in the right panel of Figure \ref{fig:astrometry}. This residual scatter is consistent with the positional uncertainties reported by \texttt{PyBDSF}, whose median values for this sample are $\sigma_{\rm RA} = 17 \pm 15$~mas and $\sigma_{\rm DEC} = 20 \pm 13$~mas.

\subsection{Computational costs}

All data processing was carried out on the Spider cluster from SURF.\footnote{\url{https://doc.spider.surfsara.nl}} The cluster nodes are equipped with AMD Rome and Bergamo architectures, offering 60 and 120 cores per node along with 960 and 1440 GB of RAM per node and local SSD storage of 10~TB and 12~TB, respectively.

The total computational cost of processing all 25 ELAIS-N1 LOFAR observations, covering a total integration time of 200~hours and a bandwidth of 53~MHz, amounted to approximately 750\,000 CPU core hours, of which about 10\% was spent on imaging. Had all observations been recorded with a 1~sec time resolution (see Section \ref{sec:data}), the processing cost would have increased by an additional \textasciitilde150\,000 CPU core hours. The total cost of processing all 25 observations is only about three times higher than that reported by \cite{sweijen2022}, who processed a single comparable dataset recorded with a 2~sec time resolution, and is close to the total computational cost reported by \cite{dejong2024a} who processed four of the LOFAR observations included in this work, of which two were recorded at 1~sec and two at 2~sec. 
These comparisons highlight the result of the computational optimisations, which are discussed in detail by \cite{dejong2025b}.

\section{Cataloguing}\label{sec:cataloguing}

We aim to generate a multi-resolution catalogue containing all available information on each radio source in our images, using all facet images at their native resolution (Appendix \ref{app:facetproperties}), combined with the deep 6\arcsec~catalogue from \cite{shimwell2025}. 
Below, we first describe how we prepare the individual catalogues at each resolution and remove unreliable low S/N detections, followed by how we combine the catalogues from different resolutions.

\subsection{Generating initial catalogues}

Initial catalogues were generated independently at each resolution (with labels 0.3\arcsec, 0.6\arcsec, and 1.2\arcsec) and for each facet, using \texttt{PyBDSF} \citep{mohan2015},\footnote{\url{https://pybdsf.readthedocs.io}} with identical parameters to what was adopted by \cite{dejong2024a}. These include a primary RMS box of 120$\times$120 pixels moved in increments of 15 pixels, supplemented by a smaller 40$\times$40 pixel box moved in increments of 10 pixels for bright sources. \`A trous wavelet decomposition was applied to break the Gaussian-fitting residual image into wavelet images at multiple spatial scales \citep{starck1994}, and a group tolerance value of 10 was adopted to enhance the merging of neighbouring Gaussians into single sources compared to the default value of 1. Other parameters were kept at their defaults. To the catalogues from our 0.3\arcsec, 0.6\arcsec, and 1.2\arcsec images, we add the 6\arcsec\ catalogue from \cite{shimwell2025}. 

We add to each catalogue a unique ID for each \textit{source}, and carry these IDs over to the Gaussian catalogue. These unique IDs are of the format \texttt{X\_Y} where the prefix \texttt{X} is the facet number for the source (see Figure \ref{fig:facet_layout}), and the suffix \texttt{Y} is the source ID from \texttt{PyBDSF}. The facet numbers are unique, while source numbers are only unique within a facet, so the combination provides a unique ID for every source in the entire field. The 6\arcsec\ catalogue came from a single image, rather than a faceted one, but we use the facet layout to create consistently formatted unique IDs (and remove sources from the catalogue which are outside of the high-resolution images).  

\subsection{Filtering low S/N sources}\label{sec:filteringSNR}

To identify and remove low S/N sources where {\ttfamily PyBDSF} may have had difficulty estimating the local noise, we recalculate the noise from the RMS maps for every source at every resolution, in a 200$\times$200 pixel box centred on the source location. This provides a better estimate for the RMS for every source. The reason we calculate the RMS for every single source using the RMS maps at every resolution is that the additional information at the other resolutions can also provide an indication of whether or not one would expect a source to be detected (e.g. if the peak intensity at 6\arcsec\ is less than 5$\sigma$ based on the noise at 1.2\arcsec , we would not expect a detection at 1.2\arcsec ). This information is not used in the S/N cuts but is included in the catalogues. For the S/N cuts we use only the noise at the relevant resolution for each source. 

We tried several combinations of criteria based on integrated flux density, peak intensity, and their associated uncertainties to filter out low-S/N sources.
While using the total flux density to define the signal for S/N keeps the most sources (90\% to 99\%, depending on the resolution), this is at the cost of having a large number of false detections or improperly fitted sources. Particularly because wavelets were used in source detection, the higher resolution images often have low S/N sources with peak intensity to noise ratio less than unity, but a total flux density to noise ratio in the hundreds. {\ttfamily PyBDSF} struggles in these cases to find the difference between the background and the FWHM wings of the source. For this reason, we chose, similar to previous high-resolution LOFAR deep field data releases \citep{sweijen2022, dejong2024a, escott2026}, to use the peak intensity to define S/N, which provides a much cleaner sample at the cost of being less complete. 

Applying a threshold requiring peak intensities to be at least 5 times the calculated local RMS, we find that 40\% of the sources are filtered out at 0.3\arcsec, 23\% at 0.6\arcsec, 27\% at 1.2\arcsec, and 20\% at 6\arcsec\ (see Table~\ref{table:snrcuts}). We also remove the corresponding Gaussian component(s) from the Gaussian catalogue. The largest removal fraction for the 0.3\arcsec~catalogue is likely due to a combination of stronger smearing effects (see Section \ref{sec:smearing}) and the resolving power, since for example some of the faint star-forming galaxies are more likely to be resolved at higher resolution (see Section \ref{sec:resolutioncompare}). 
Given the substantial source filtering, and for reproducibility, we also release the `raw' \texttt{PyBDSF} outputs in case someone wishes to look up individual faint sources. However, we advise caution, as these catalogues include false detections and are unreliable for large population studies.
Future work will explore how to improve the source detections across resolutions.

\begin{table}
\caption{Number of Gaussians and sources before and after removing sources.}
\begin{tabular}{lcccc}
  & 0.3\arcsec & 0.6\arcsec & 1.2\arcsec & 6.0\arcsec\\ \hline
Gaussians       & 58\,261 & 44\,106 & 39\,820 & 71\,412 \\ 
S/N $>$ 5   & 35\,842 & 34\,168 & 29\,943 & 58\,821 \\ \hline 
Sources         & 51\,779 & 39\,413 & 35\,079 & 59\,993 \\ 
S/N $>$ 5       & 30\,965 & 30\,174 & 25\,656 & 47\,882 \\ 
Single          & 28\,484 & 28\,585 & 23\,811 & 41\,219 \\ 
Multiple        & 2\,481 & 1\,589 & 1\,845 & 6\,663 \\ \hline
\end{tabular}
\tablefoot{`Single' refers to single-component sources and `multiple' to multiple component sources, according to \texttt{PyBDSF}.}
\label{table:snrcuts}
\end{table}

Quantifying completeness requires separate simulations at every resolution (due to differences in surface brightness sensitivity and the wide range of source structure at different resolutions), which are beyond the scope of this paper. Such simulations would also need to account for time and bandwidth smearing, which is another factor that affects the completeness for the 0.3\arcsec\ and 0.6\arcsec\ images, given our peak-based S/N threshold (see Section~\ref{sec:smearing}). Simulations would also provide the false positive rate, but the fact that we aim to match across different resolutions already helps mitigate this. For example, a marginal detection at 0.3\arcsec\ is likely a real source if it is also detected at 6\arcsec. Future work aiming to quantify the true false-detection rate should therefore focus on the unmatched sources at each resolution, and fold in \texttt{PyBDSF}'s wavelet decomposition.

\begin{figure}
    \centering
    \includegraphics[width=0.6\linewidth,clip,trim=20cm 0cm 0cm 2cm]{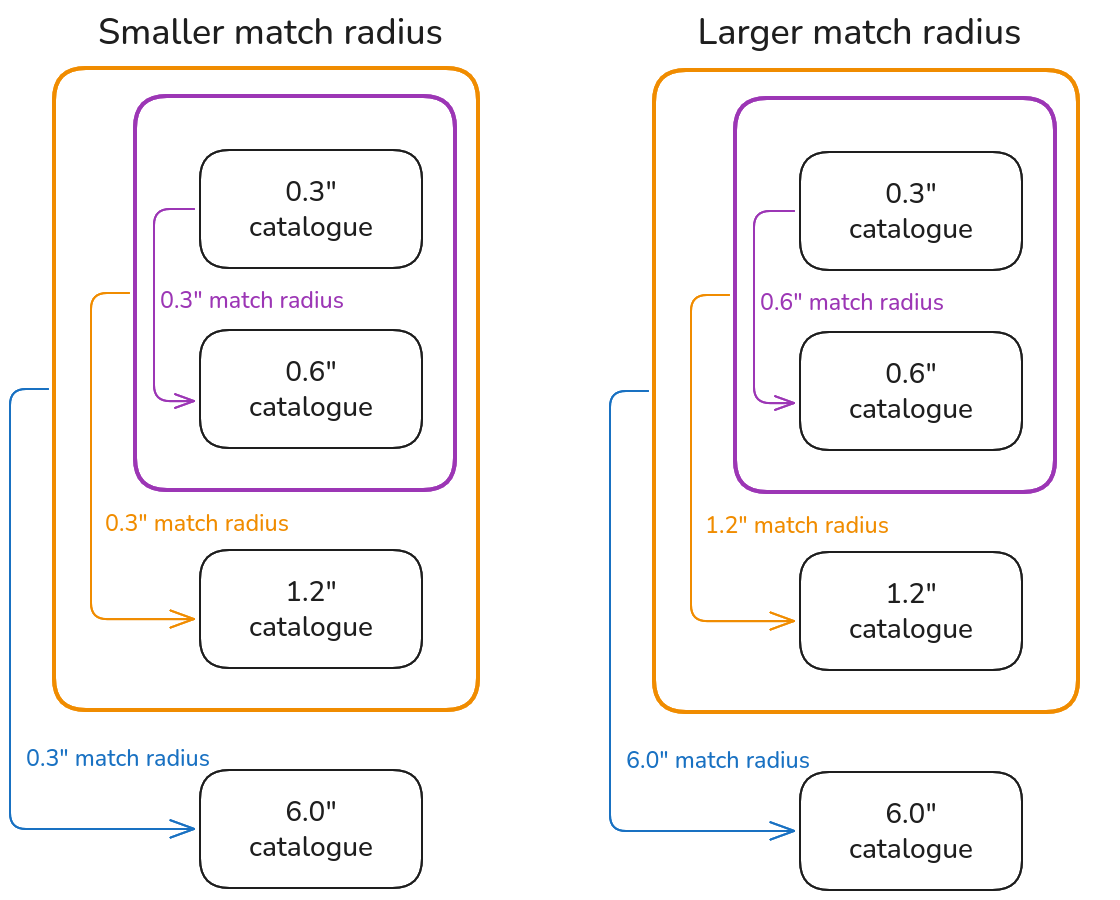}
    \caption{Schematic diagram of how we change the match radius across different resolutions. This example is only for the 0.3\arcsec\ catalogue, but we repeated this process starting from the 0.6\arcsec\ and 1.2\arcsec\ to account for all possible combination of resolutions. }
    \label{fig:matching}
\end{figure}

\subsection{Cross-matching resolutions}

Next, we use positional matching based on the RA and DEC in the catalogues from each resolution. This requires setting a match radius. Typically the FWHM of the beam can be used, but when matching two different resolutions, there are two different FWHMs. After checking that indeed the larger FWHM contains all possible matches using the smaller FWHM, we use only the larger match radius. Figure~\ref{fig:matching} shows a schematic diagram which helps visualise this process. We first use the 0.3\arcsec\ catalogue as a reference, and then repeat the process from 0.6\arcsec\  (matching to 1.2\arcsec\ and 6.0\arcsec ), and repeat again starting from 1.2\arcsec\ (matching to 6.0\arcsec ). All matches are kept, as are sources where there are no matches at other resolutions. These three resulting catalogues are merged and we remove duplicates and rows whose detections are all contained in another row with detections at more resolutions. These are matches where a unique single detection, unique pair, or unique triple match exists on a different row with more information. 

In the multi-resolution catalogue, each row contains a unique combination of information across the four resolutions, and can have information at 1, 2, 3, or 4 resolutions, with null values for any resolution where there is not a detection. The differing point source and surface brightness sensitivities at different resolutions, combined with source structure, means that not every source is detected at every resolution. 
It is important to note that in the final catalogue, one \textit{row} does not equal one \textit{source}. For example, if a source has a single component at 6\arcsec\ but breaks into two components at 0.3\arcsec, there will be two rows, one for each of the individual 0.3\arcsec\ components, each carrying the information from the associated component at 6\arcsec\ resolution. The complete set of information on this \textit{source} therefore requires selecting both of these rows to get the information on the individual components. 

We identified 1\,115 locations for visual inspection. These are locations where sources with a single detection at a lower resolution break into multiple components at higher resolutions; we recorded a single common position (that from the lowest resolution) for inspection. These will be explored in future work. 

\begin{table}
\caption{Number of rows and unique sources for each resolution in the final multi-resolution catalogue.}
\begin{tabular}{ccc}
 & No. rows & No. sources \\ \hline 
0.3\arcsec & 30\,978 & 30\,965 \\ 
0.6\arcsec & 30\,355 & 30\,174 \\ 
1.2\arcsec & 26\,239 & 25\,656 \\ 
6.0\arcsec & 48\,762 & 47\,882 \\ 
 \hline 
Combined & 63\,841 & 62\,916 \\ \hline
\end{tabular}
\tablefoot{The number of rows exceeds the number of unique sources because a source that \texttt{PyBDSF} detected as multiple separate components at a certain resolution but as a single component at another resolution will occupy multiple rows despite being one unique source.}\label{table:final}
\end{table}

\begin{figure}
    \centering
    \includegraphics[width=1.0\linewidth]{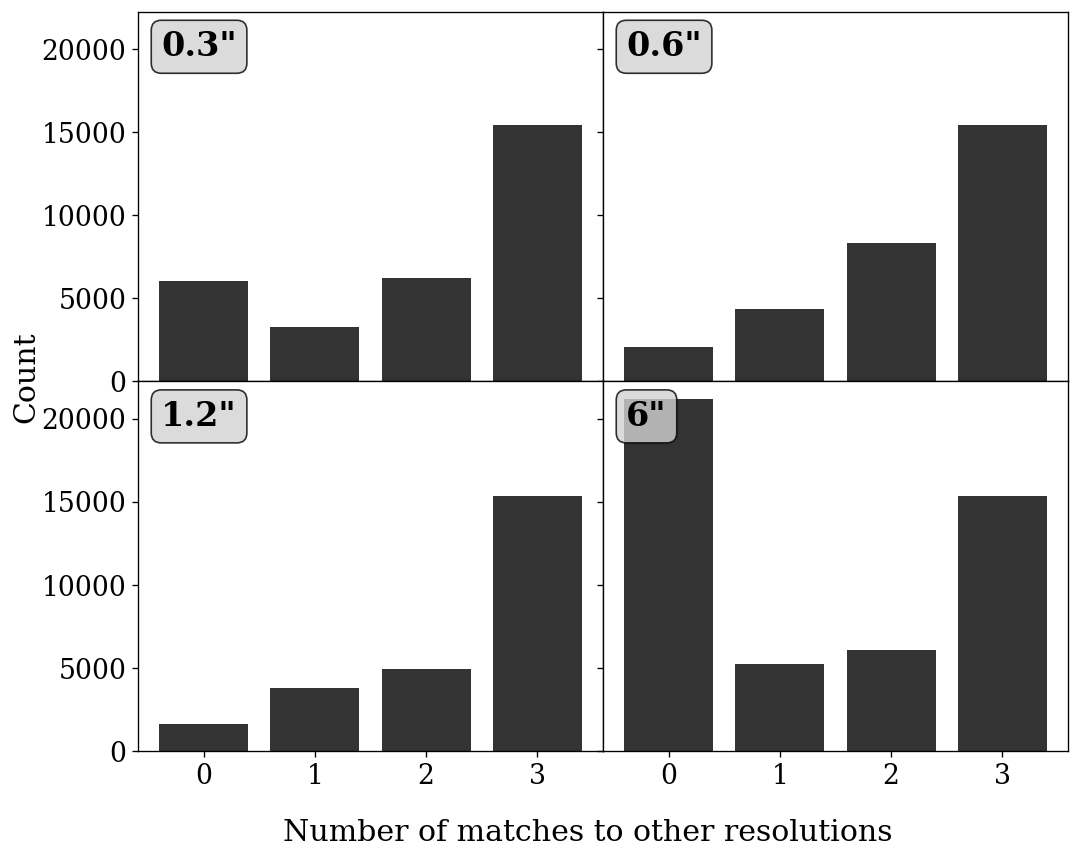}
    \caption{For each resolution, the number of other resolutions to which sources are matched. }
    \label{fig:nmatches}
\end{figure}

\subsection{Cataloguing summary}

To summarise, the final combined catalogue has 
63\,841 rows with 62\,916 unique sources. Table~\ref{table:final} summarises the number of rows and unique source IDs at each resolution. The source density is determined in the FoV ($2.5\times2.5$~deg$^{2}$) to be 4\,954, 4\,828, 4\,105, and 7\,661 per square degree for 0.3\arcsec , 0.6\arcsec , 1.2\arcsec , and 6.0\arcsec , respectively. The 0.3\arcsec\ catalogue alone has, to our knowledge, the highest source density of any survey with sufficient resolution to identify AGN via their brightness temperature \citep{morabito2022a}. 

\begin{figure}
    \centering
    \includegraphics[width=1.0\linewidth]{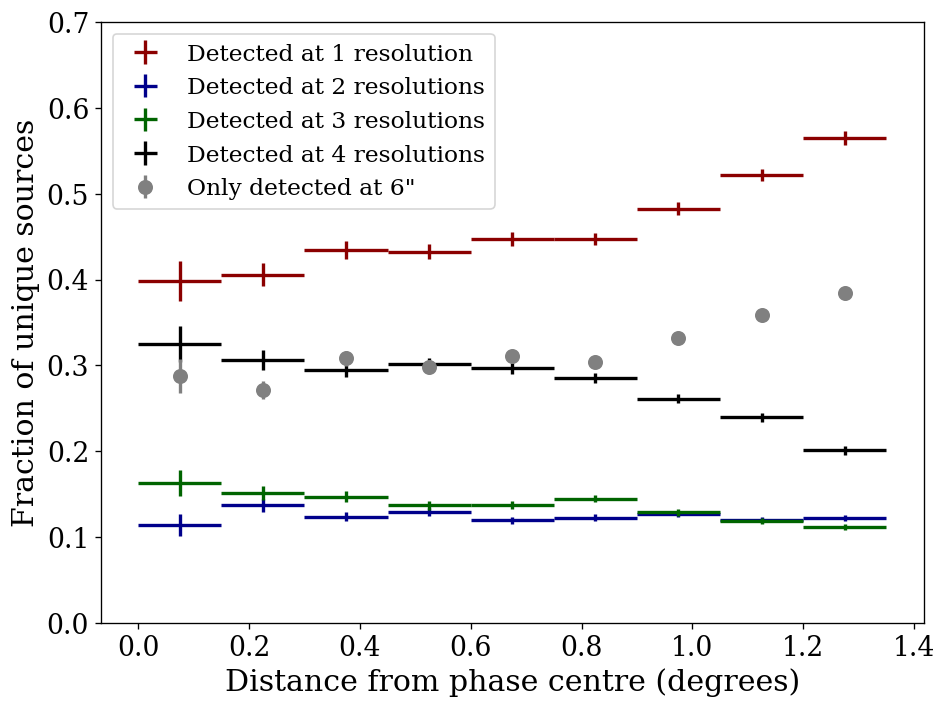}
\caption{Fraction of unique sources detected at one, two, three, or four of the angular resolutions (labelled with 0.3\arcsec, 0.6\arcsec, 1.2\arcsec, and 6\arcsec) as a function of distance from the phase centre. Grey points show the fraction of sources detected only at 6\arcsec. Horizontal error bars indicate the bin widths, and vertical error bars indicate the Poisson uncertainties.}
    \label{fig:nres_dist}
\end{figure}

We show in Figure~\ref{fig:nmatches} the number of matches at other resolutions, for each resolution. Of the 62\,916 unique sources, 50.5\% are detected at more than one resolution and 25.5\% at all four resolutions, which highlights the benefit of combining multiple resolutions. The 1.2\arcsec\ images have the lowest number of unique detections, as this resolution has the lowest point-source sensitivity and is a balance between the highest and lowest angular resolution in terms of surface brightness sensitivity. Figure~\ref{fig:nres_dist} shows the fraction of unique sources as a function of distance from the phase centre for sources detected at 1, 2, 3, or 4 resolutions with the fraction of sources detected only at 6\arcsec\ for comparison. Within the inner 0.3~deg, about 40\% of the sources are detected at a single resolution, mostly at 6\arcsec\ only, indicating that most single-resolution detections are faint or extended sources below the detection threshold at higher resolutions. Beyond $\sim$0.8~deg, the fraction of sources detected at all four resolutions decreases, while the fraction detected only at 6\arcsec\ increases, consistent with the increasing time and bandwidth smearing (see Section~\ref{sec:smearing}) and the narrower primary beam of the international stations. The fractions of sources detected at two or three resolutions remain roughly constant with distance, as sources that lose for example their 0.3\arcsec\ detection towards the edge of the field move into these categories, while others move from them to the single resolution category.

\begin{figure*}
\centering
\includegraphics[width=0.99\linewidth]{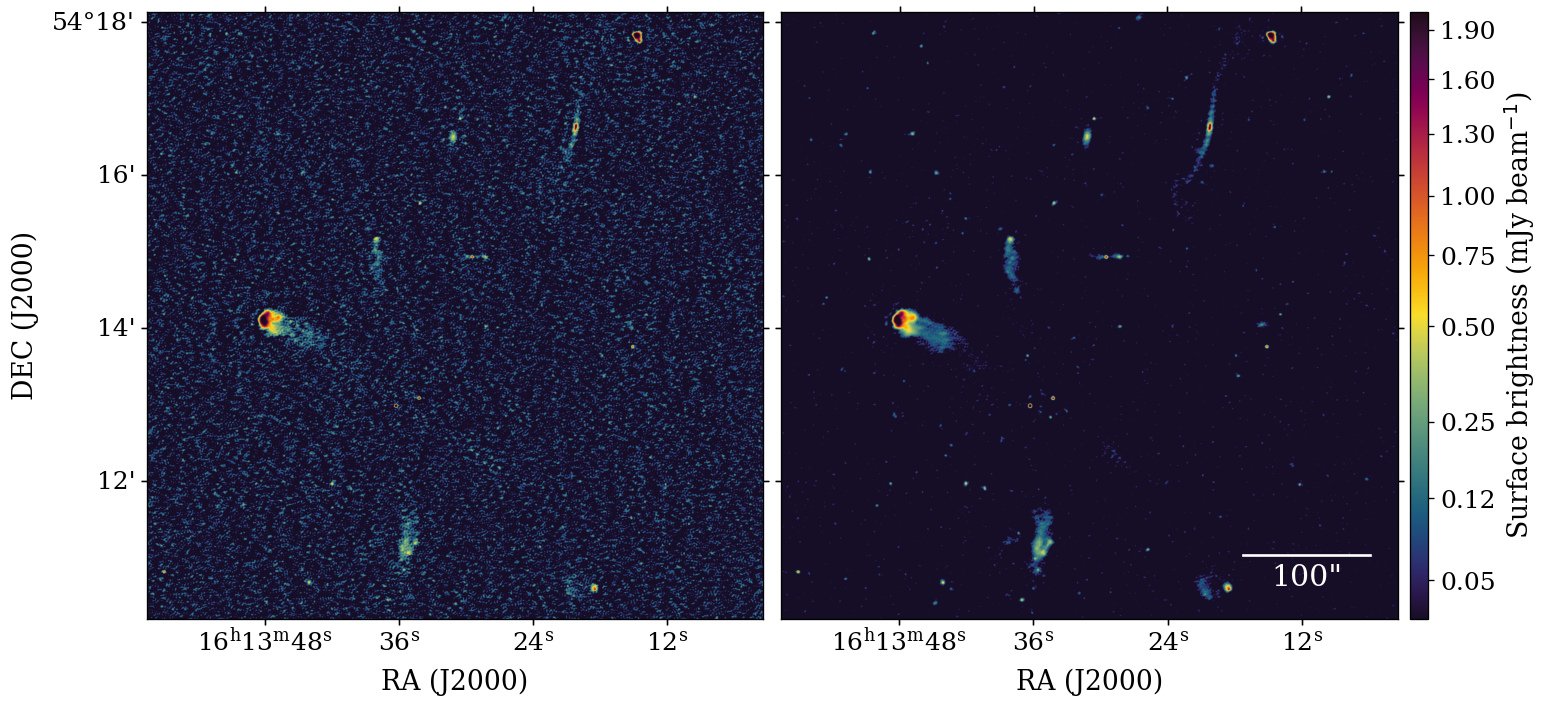}
\includegraphics[width=0.99\linewidth]{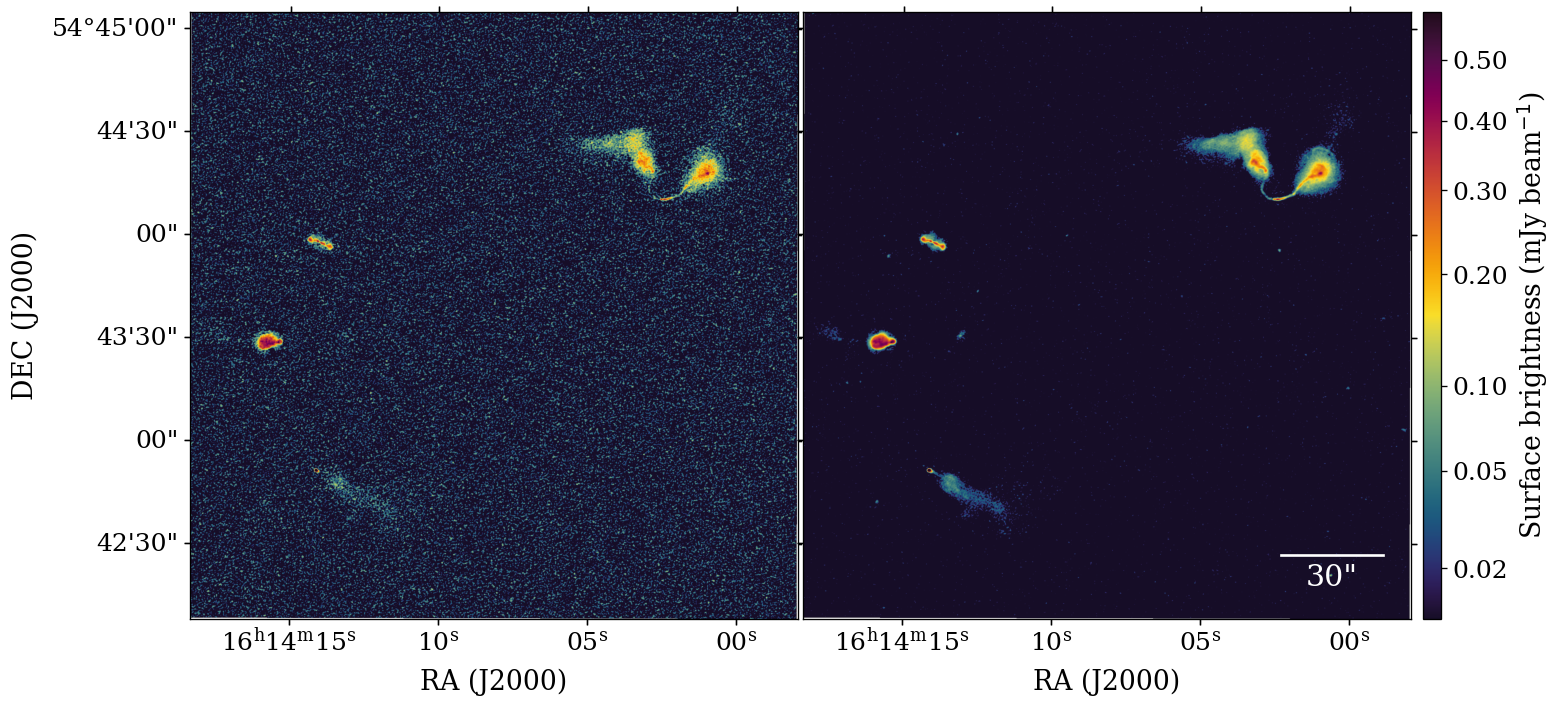}
  \caption{Comparison between imaging a single LOFAR observation (\textit{left}) and 25 observations (\textit{right}) for two different sky areas. The top images have been restored to a circular beam of 1.5\arcsec, the bottom images have been restored to a circular beam of 0.4\arcsec. The images in the left panels were created with the observation with ID \texttt{L686962}. The colour map was generated with \texttt{CMasher} \citep{cmasher}.}
\label{fig:fireworks}
\end{figure*}

\section{Data quality}\label{sec:results}

With our images and catalogues, we examine in this section several aspects that determine the quality of the data. These include the depth, smearing, confusion noise, and polarisation.

\begin{figure}
 \centering
\includegraphics[width=0.98\linewidth]{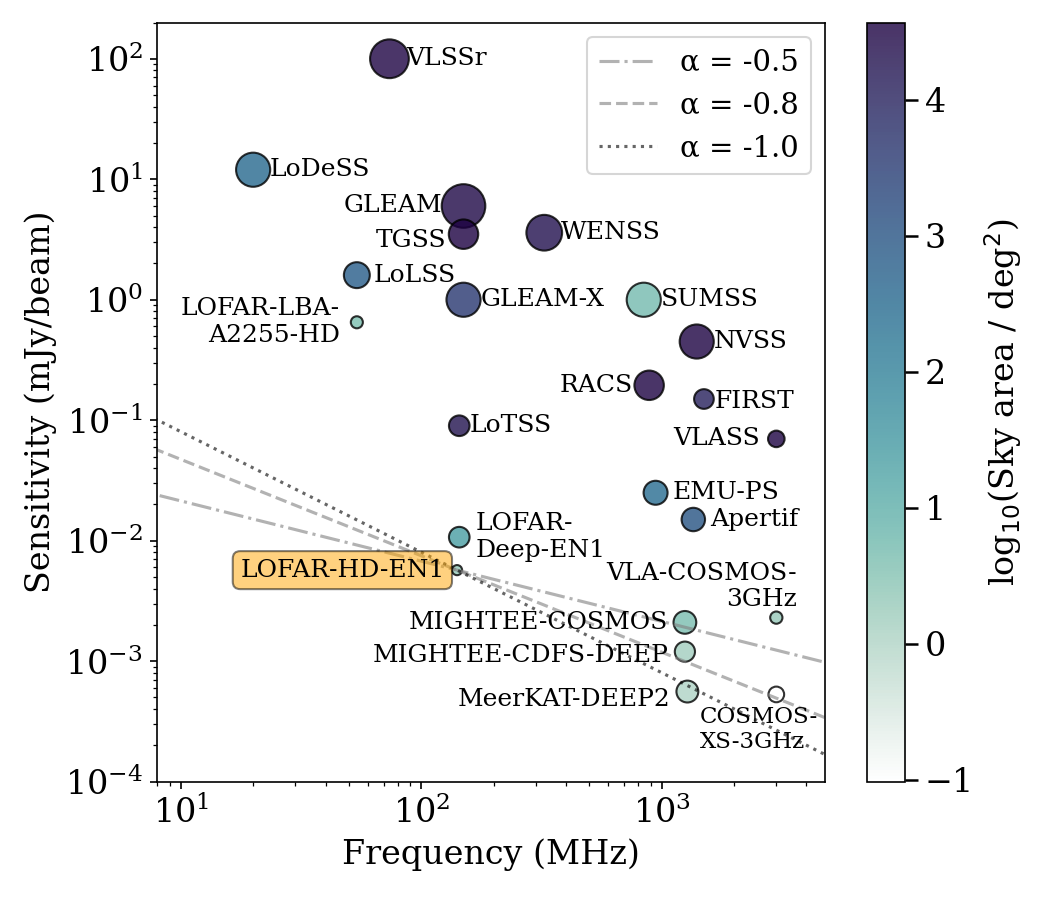}
  \caption{Sensitivity as a function of frequency for various surveys and deep images. The radius of each circle represents the square root of the beam size, and the colour represents the survey area. `LOFAR-HD-EN1' corresponds to the deepest image presented in this paper, where we show the area-weighted RMS noise within the inner 1.5$\times$1.5\,deg$^2$ of the field (5.8\,$\upmu$Jy\,beam$^{-1}$) to highlight the central depth in our highest resolution images. From the MIGHTEE images released by \cite{hale2025}, the deepest CDFS-DEEP and Cosmic Evolution Survey (COSMOS) images are displayed. The dashed lines show the sensitivity of our image extrapolated to other frequencies using different spectral indices: $\alpha=-1$ for steep-spectrum sources, $\alpha=-0.8$ for a typical mix of LOFAR-detected radio sources according to \cite{bohme2023}, and $\alpha=-0.5$ as measured for star-forming galaxies in ELAIS-N1 \citep{an2024}.}
\label{fig:surveys_depth}
\end{figure}

\subsection{Depth}

We achieve a central point-source sensitivity of 5.3\,$\upmu$Jy\,beam$^{-1}$ in facet 2 at 0.33\arcsec$\times0.39$\arcsec\ and an area-weighted RMS noise of 7.4\,$\upmu$Jy\,beam$^{-1}$ over the full 2.5$\times$2.5\,deg$^2$ field at 0.36\arcsec$\times0.45$\arcsec\ resolution (Appendix \label{app:facetproperties}). The central point-source sensitivity is a factor of two lower than that of the central point-source sensitivity from the 6\arcsec\ wide-field image by \cite{shimwell2025}, who imaged 505~hours of LOFAR data. The improvement is due to the larger collecting area and higher sensitivity of the international LOFAR stations, as well as the fact that our images are not affected by confusion noise at sub-arcsecond~resolution (see Section \ref{sec:confusion}), unlike imaging at 6\arcsec~with LOFAR \citep[e.g.][]{sabater2021, shimwell2025}. Compared to \cite{dejong2024a}, who imaged 32~hours of the same field at the same resolution, we achieve a factor of 2.6 improvement. This is consistent with the expected square-root scaling with total integration time. To compare the sensitivity improvements between imaging with a standard single 8-hour LOFAR observation and our 200-hour imaging, we show in Figure~\ref{fig:fireworks} images of the same sky area at both corresponding total integration times at two different resolutions. The deeper images in the right panels recover more diffuse emission from jets and reveals a larger number of faint background point sources, mainly corresponding to distant star-forming galaxies \citep[e.g.][]{best2023}.

In Figure \ref{fig:surveys_depth} we compare the point-source sensitivity from our deepest high-resolution image with other surveys and deep images \citep{condon1998, sumss1999, schinnerer2007, lane2014, helfand2015, wayth2015, smolcic2017, intema2017, shimwell2017,shimwell2022,shimwell2025,shimwell2026, mauch2020, lacy2020, algera2020, vlugt2021, norris2021, sabater2021, hurleywalker2022, adams2022, gasperin2023, groeneveld2024, duchesne2023, Duchesne2025, hale2025}. 
As most of the deepest images cover a considerably smaller FoV, we use for this comparison the area-weighted RMS noise within the inner 1.5$\times$1.5\,deg$^2$ of our image, which is 5.8\,$\upmu$Jy\,beam$^{-1}$ and highlights better the significant depth we reach in this work.
We find in this figure that our image is the most sensitive radio image below 1.28~GHz. The comparison with higher frequency images depends on the assumed spectral index ($\alpha$, where $S_{\nu}\propto\nu^{\alpha}$). For a steep index of $\alpha=-1$, our image reaches an equivalent point-source sensitivity comparable to that of the MeerKAT DEEP2 image \citep{mauch2020}, whereas for $\alpha \sim -0.8$, typical of synchrotron-dominated radio sources \citep{bohme2023}, the MeerKAT DEEP2 image remains deeper, while the image from the COSMOS-XS 3~GHz survey \citep{algera2020, vlugt2021} reaches a more comparable equivalent point-source sensitivity. For $\alpha \sim -0.5$, which \citet{an2024} found to be typical of star-forming galaxies between 150 and 610~MHz in ELAIS-N1, both of these images and the Chandra Deep Field South (CDFS) image from the MeerKAT International GHz Tiered Extragalactic Exploration Survey \citep[MIGHTEE;][]{hale2025} all reach a better equivalent sensitivity than our image.

When we compare the deepest observations below an equivalent sensitivity at 150~MHz of 20~$\upmu$Jy~beam$^{-1}$, we find the highest number of detections in the LOFAR-Deep-EN1 from \cite{shimwell2025}, owing to its FoV being four times larger than ours. However, since our multi-resolution catalogue is combined with theirs for the central $2.5\times2.5$~deg$^{2}$ of their FoV, our catalogue has a higher source density. Our FoV is significantly larger than those of MeerKAT DEEP2, COSMOS-XS 3~GHz, and MIGHTEE CDFS-DEEP, yielding higher numbers of source detections than theirs, although our point-source sensitivity drops fast towards the edge of the field. This benefits science dependent on larger number statistics. In addition, our 0.3\arcsec\ images far surpass the resolution of every image and survey shown in Figure~\ref{fig:surveys_depth}, adding scientific value that goes beyond depth alone \citep{morabito2025a}.

\begin{figure*}
 \centering
\includegraphics[width=0.33\linewidth]{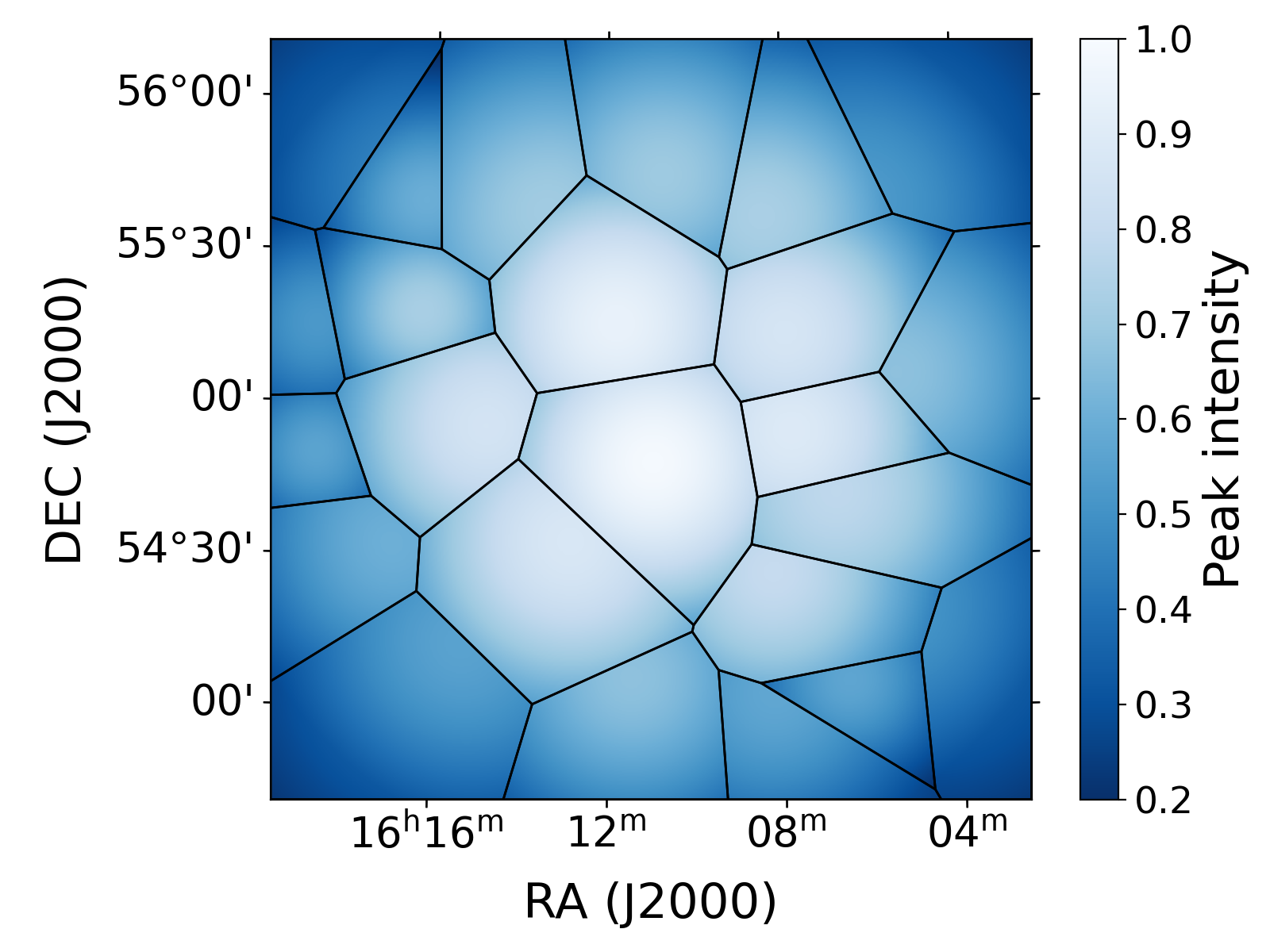}
\includegraphics[width=0.33\linewidth]{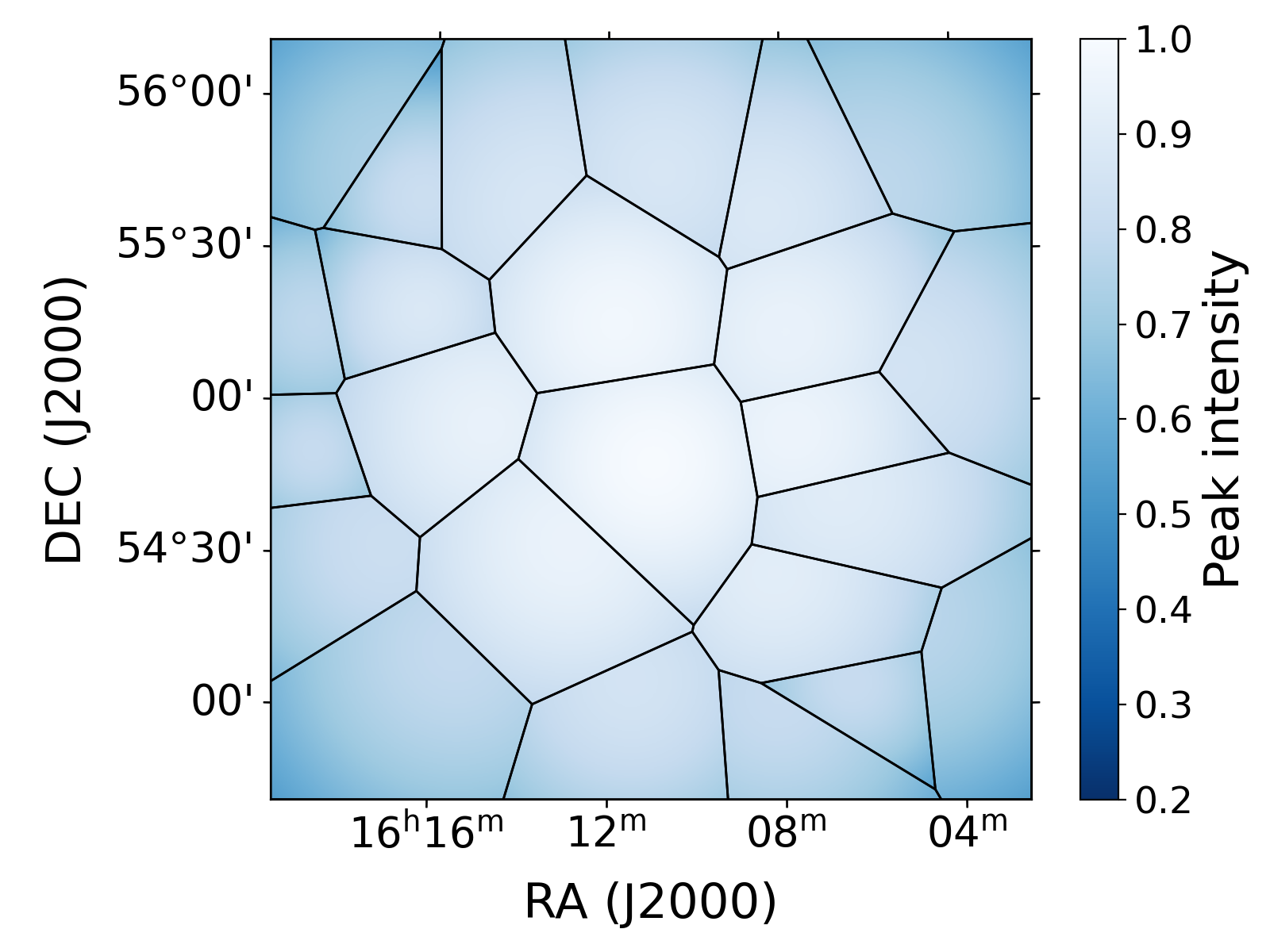}
\includegraphics[width=0.33\linewidth]{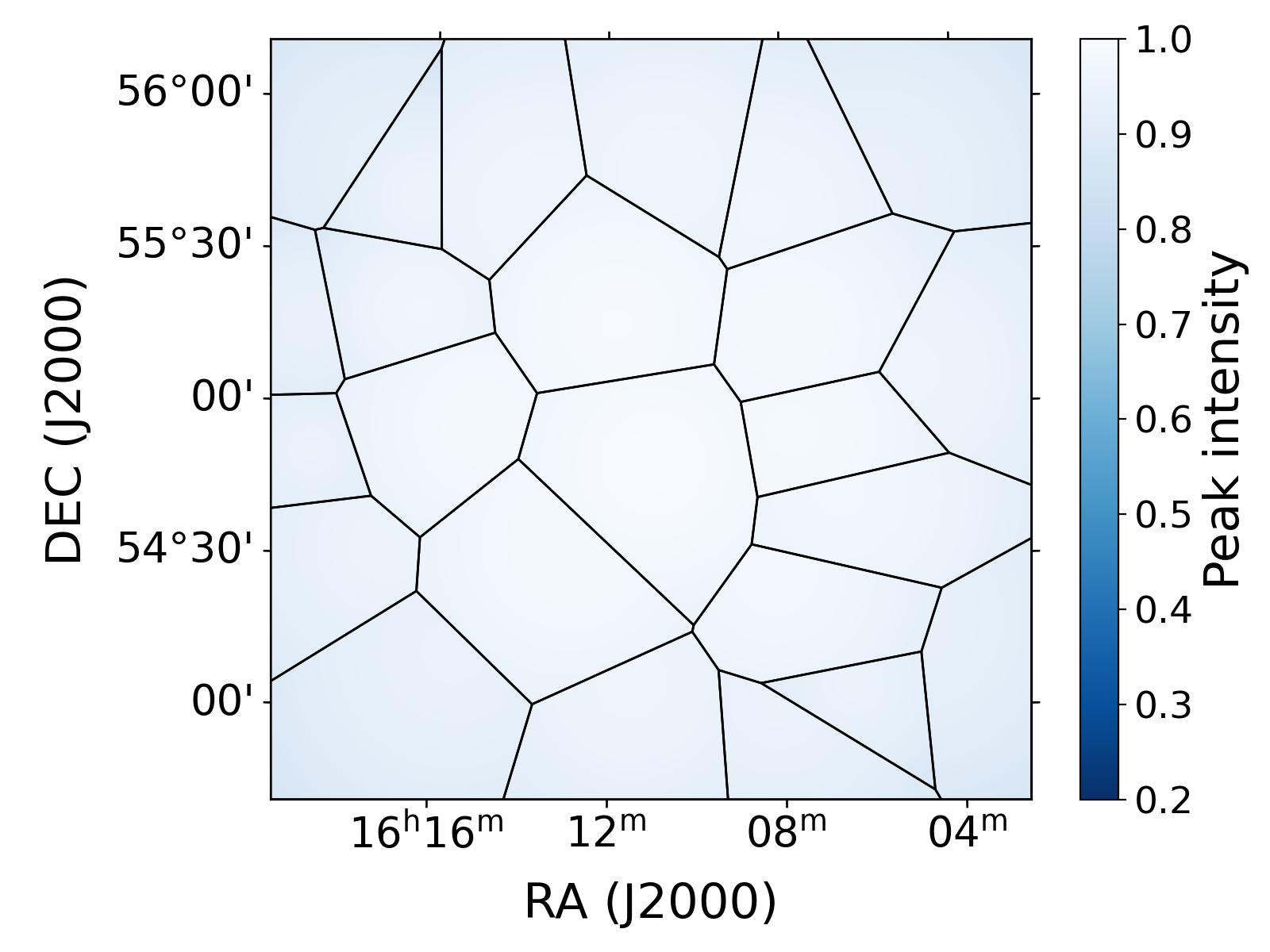}
  \caption{Leftover peak intensities in our final wide-field images due to time and bandwidth smearing across the FoV, according to the formulas from \cite{bridle1999}, for each mosaic at their restored resolution: $0.36\arcsec\times0.45\arcsec$ (\textit{left}), $0.58\arcsec\times0.62\arcsec$ (\textit{centre}), and $1.0\arcsec\times1.5\arcsec$ (\textit{right}). These smearing maps take into account the averaging after facet-subtraction and capture the weighted combination of observations recorded at 1~sec and 2~sec (Section~\ref{sec:data}).}
\label{fig:smearing}
\end{figure*}

\subsection{Smearing}\label{sec:smearing}

Bandwidth and time smearing are key factors that degrade image quality and source detections. Bandwidth smearing causes sources to broaden radially, while time smearing produces blurring along the tangential direction relative to the phase centre. This results in a decrease in peak intensity. Additional smearing may also be introduced by ionospheric effects \citep{kassim2007, loi2015}, even after applying DD corrections. Furthermore, SVA can introduce additional smearing at the level of a few percent \citep{dejong2025}. Although smearing conserves the integrated flux density of a source, the accompanying loss of peak intensity means that part of the emission from faint sources near the detection threshold falls below the noise, so their integrated flux densities are systematically underestimated. Since we have removed sources from our catalogue with low S/N values based on their peak intensities (see Section \ref{sec:filteringSNR}), smearing may also affect the completeness of our catalogue. Below we assess the various smearing effects on our data.

According to the formulas of \citet{bridle1999}, for LOFAR datasets typically stored in the LTA with a time resolution of 1~sec and a channel width of 12.21~kHz, the peak intensity of a point source is already reduced by about a factor of two at 0.3\arcsec~resolution, 1.25~degrees from the phase centre. On top of this, as described in Section~\ref{sec:data}, 11 of the 25 observations that we used to create our wide-field image were recorded with a 2~sec time resolution. As a result, the total time and bandwidth smearing in our images becomes significant towards the outer edge of the field \citep[e.g.][]{sweijen2022, dejong2024a}.
In addition to the intrinsic smearing, we also apply further averaging to the datasets from each facet prior to imaging \citep{dejong2025b}. These averaging factors are calculated under the assumption that the already present smearing at the edge of a 2.5$\times$2.5~deg$^{2}$ LOFAR FoV at 1~sec and 12.21~kHz is acceptable, as that keeps the data volumes manageable.\footnote{Without any averaging, the data would require approximately 1.5~PB of storage.} Hence, a facet that fits in a 0.8$\times$0.8~deg$^{2}$ box will be averaged by a factor of three in time and frequency. This means that the bandwidth and time smearing from the full dataset is carried over towards the edges of the facets after averaging. 

To illustrate the impact of time and bandwidth smearing effects across the field, Figure~\ref{fig:smearing} shows the estimated peak intensity reduction for baselines corresponding to the three angular resolutions (see Section~\ref{sec:imaging}). We find that, as expected, the 0.3\arcsec\ images suffer the most from smearing, with peak intensity losses of up to 80\% at the edge of the field, whereas the lower resolution images are less affected, as time and bandwidth smearing scale with angular resolution. We can also see the additionally introduced smearing at the edges of each facet. As an alternative check of the effects of time and bandwidth smearing, \citet{dejong2024a} measured the ratio of peak intensity to integrated flux density for compact sources as a function of distance from the phase centre, using a subset of the same ELAIS-N1 observations with a similar mixture of 1~sec and 2~sec data. The trends they found agree with our predictions in Figure~\ref{fig:smearing}. However, these figures should be interpreted with caution as the S/N threshold biases the source selection and sources are more likely to be resolved at higher resolution, which lowers their peak intensities independently of smearing (see Section \ref{sec:resolutioncompare}). In Appendix~\ref{appendix:smearing} we briefly discuss how finer time and frequency resolution in the initial data reduces time and bandwidth smearing, at increased computational cost.

To estimate the impact of ionospheric smearing, we re-imaged the most central facet (facet~2) and a facet at the edge of the field (facet~10) for one observation corresponding to calm ionospheric conditions (\texttt{L769407}) and one under more disturbed ionospheric conditions (\texttt{L686948}). We selected these observations by visually inspecting the self-calibration images and calibration solutions from all our processed observations. We then selected compact sources (labelled as single components by \texttt{PyBDSF}) with peak intensities exceeding ten times the local RMS noise and used the ratio of peak intensity to integrated flux density as a measure of smearing. For the observation with the more disturbed ionosphere, we find that this ratio is on average $15\pm5$\% lower, indicating that ionospheric conditions do introduce peak intensity losses, though at a lower level than the intrinsic time and bandwidth smearing away from the phase centre for observations stored at 1~sec and 12.21~kHz and imaged at 0.3\arcsec.

We note that SVA can also introduce smearing effects due to precession, nutation, and annual aberration, particularly since our observations span a three-year interval (affecting precession and nutation terms) and were taken at different times of the year (introducing annual aberration effects). As mentioned in Section \ref{sec:imaging}, we therefore oversampled the time axis by a factor of two, which reduces these effects to only a few percent at the edge of the field \citep{dejong2025}. This contribution is negligible compared to the much larger smearing caused by the time and frequency averaging mentioned above. 

\subsection{Confusion noise}\label{sec:confusion}

Confusion noise arises when faint background radio sources close to or below the detection threshold become so numerous that across the FoV multiple sources fall within a single synthesised beam. In this regime, each faint background source contributes a scaled PSF, and the superposition of many such contributions adds a substantial non-Gaussian noise-like background in addition to the Gaussian thermal noise. This introduces limits to the obtained sensitivity, which is set by the angular resolution and the underlying source counts \citep{condon1974, condon2012, zwart2015, mauch2020}. Confusion noise is known to be a dominant limitation for deep low-resolution wide-field imaging. In particular, LOFAR observations using only the Dutch stations start to be dominated by confusion noise when imaging with several hundred hours of data, such that further increases in observing time yield only marginal sensitivity improvements \citep[e.g.][]{sabater2021, shimwell2025}.

Since the confusion noise scales approximately as the square of the beam size \citep{condon1974},
an image at 0.3\arcsec\ resolution can be made about 400 times deeper than one
at 6\arcsec\ before confusion becomes limiting. 
We adopt 0.3\arcsec\ as a representative resolution for simplicity, although the restoring beams of our highest-resolution facet images are larger (see Appendix~\ref{app:facetproperties}). Note that at 144\,MHz, LOFAR can reach a resolution of up to ${\sim}0.2$\arcsec\ with its longest baselines \citep{morabito2022}, and the addition of more distant international stations with LOFAR2.0 may push this further \citep[e.g.][]{dechev2023}, which would further reduce the confusion noise.
\citet{shimwell2025} reached in their 6\arcsec~image a central RMS of $10.7\,\upmu$Jy\,beam$^{-1}$, against a thermal noise of $7.5\,\upmu$Jy\,beam$^{-1}$, implying a confusion noise contribution of
$\sqrt{10.7^2-7.5^2}=7.6$\,$\upmu$Jy\,beam$^{-1}$. Scaled to 0.3\arcsec, confusion noise would therefore only become relevant when imaging close to a thermal noise of 20\,nJy\,beam$^{-1}$. When accounting for the higher point-source sensitivity of the international LOFAR stations, reaching that thermal noise would require on the order of 20 million hours of observing time. This total integration time extends well beyond the operational lifetime of LOFAR, even if the point-source sensitivity were an order of magnitude better. Hence, further extensions of this work to even deeper imaging (see Section~\ref{sec:future}) will not be constrained by confusion noise. However, other effects, such as sidelobe confusion from sources outside the FoV, residual DD calibration errors, and dynamic range limitations around bright sources, are likely to limit the achievable depth before the thermal noise limit is reached.

\subsection{Polarisation}

As described in Section~\ref{sec:calibration}, in step 3 of the calibration process we corrected for polarisation leakage and for differences between the right- and left-handed circular polarisations using an unpolarised calibrator. In step 9, we aligned the polarisation angles across the different observations using a known linearly polarised source (7C~1604+5447), thereby enabling deep polarisation imaging.
This follows the same procedure described in \cite{vanweeren2026}. Since these are the deepest data for polarisation studies below 200~MHz, we briefly discuss the quality and refer to future work for more in-depth analysis.

\begin{figure*}
 \centering
 \includegraphics[width=0.25\linewidth]{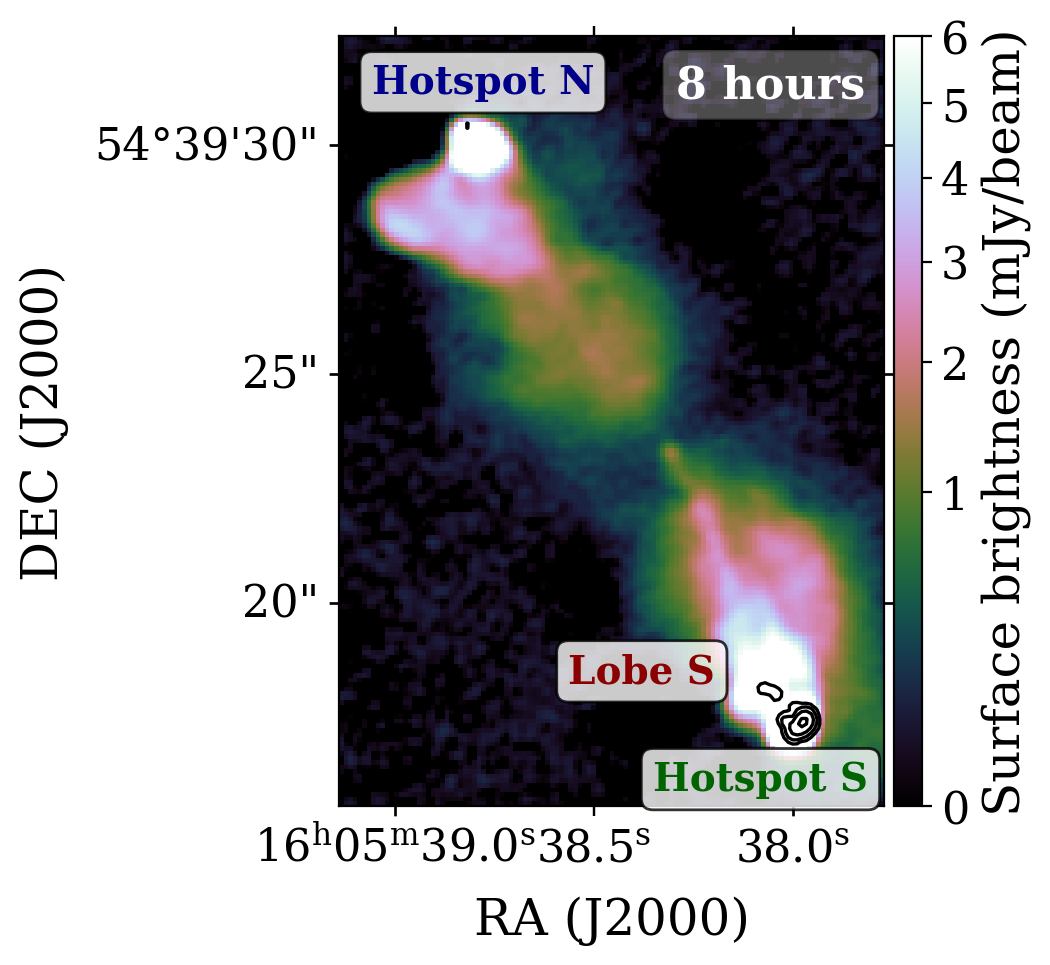}
\includegraphics[width=0.23\linewidth]{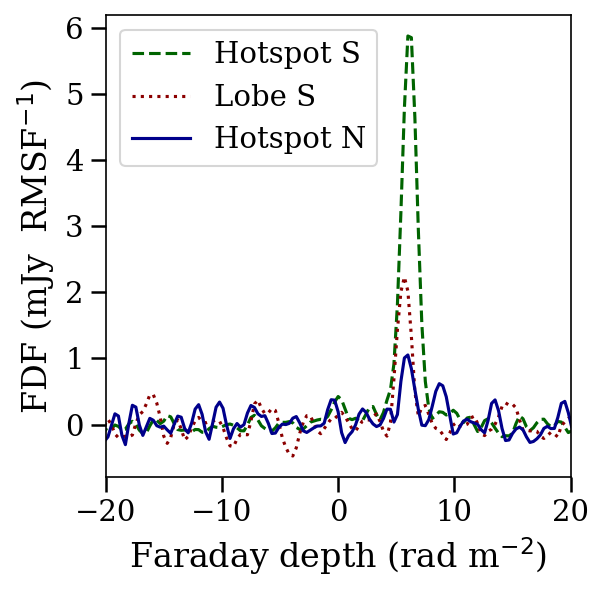}
\includegraphics[width=0.25\linewidth]{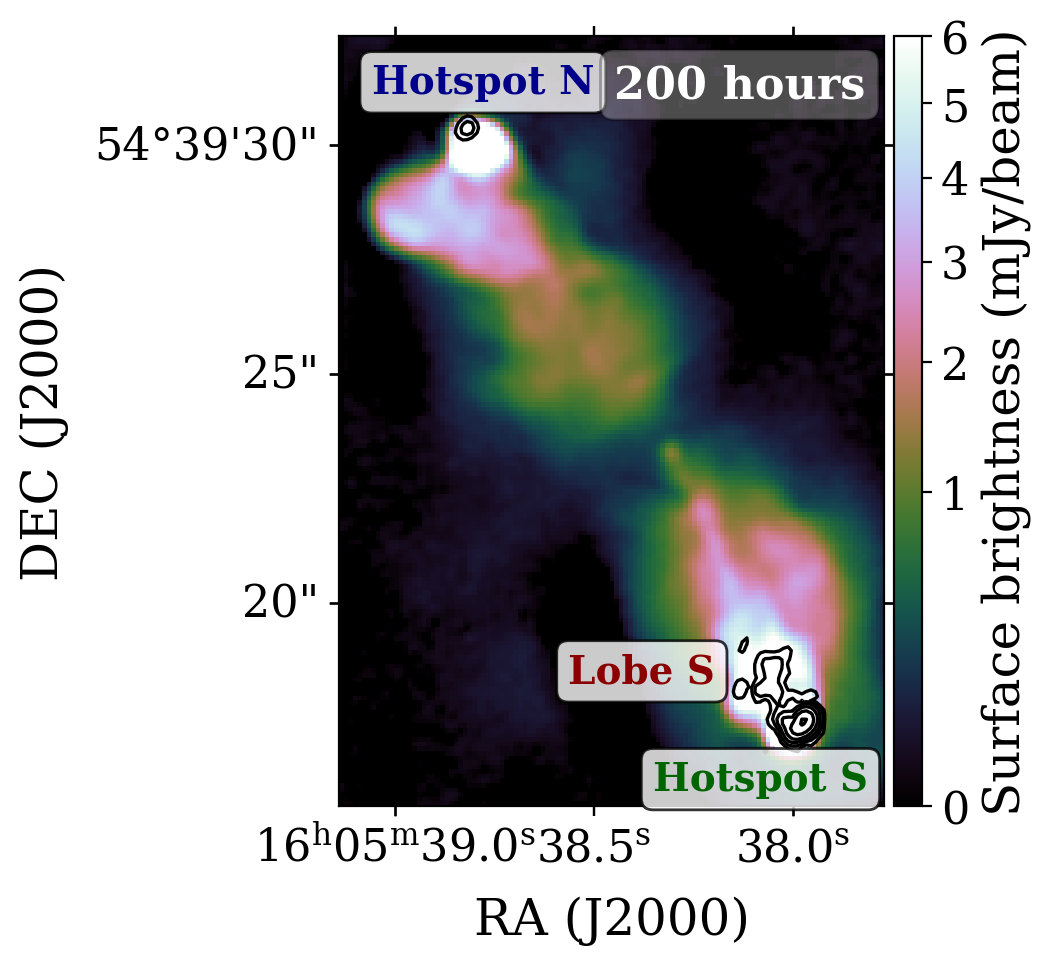}
\includegraphics[width=0.23\linewidth]{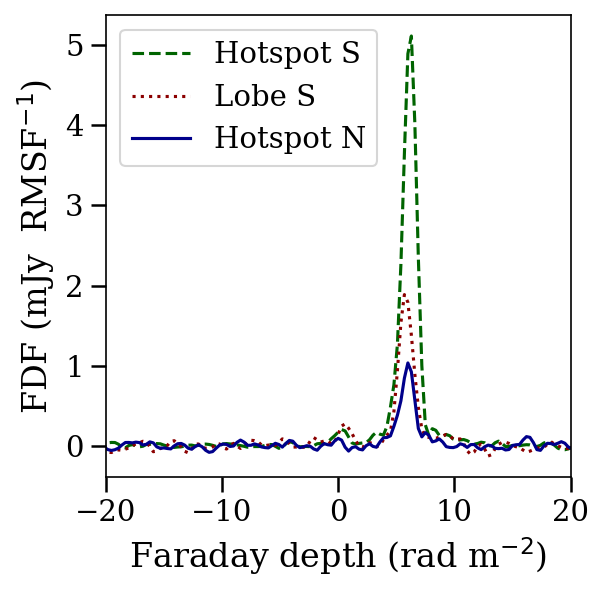}
  \caption{Stokes I images of the polarisation calibrator (7C~1604+5447) with FDF spectra at 0.3\arcsec~resolution shown with 8 and 200 hours of LOFAR data and with polarised intensity contours overlaid at levels of $10\times\sigma\times(1,2,4,8,...)$.  Following \cite{vanweeren2026} and for visualisation purposes, the spectra of `Hotspot N' and `Lobe S' were scaled by factors of five and two, respectively.}
\label{fig:polarisation}
\end{figure*}

After alignment of the different observations we find for our leakage calibrator source (ICRF\,J160607.6+552135) an instrumental leakage of 0.12\% for linear polarisation (Q,U) and 0.02\% for circular polarisation (V), which is approximately a factor of two better than the typical leakage reported for the LOFAR Two-metre Sky Survey (LoTSS) at 6\arcsec\ resolution \citep{shimwell2022}. To demonstrate deep polarisation imaging, we performed RM-synthesis \citep{brentjens2005} with the \texttt{RM-Tools} package \citep{purcell2020,vaneck2026} on Stokes Q and U images from our polarisation calibrator, using 8~hours and 200~hours of LOFAR data, and following the same imaging settings as \cite{vanweeren2026}. This gives us the Faraday dispersion functions (FDF) at the different linearly polarised regions as illustrated in Figure~\ref{fig:polarisation}. We use contour levels of 10$\sigma$ to stay above the false-detection level of peaks in the polarised intensity cube across the Faraday depth range searched. This figure show that the FDF is considerably less noisy in the 200-hour dataset than in the single 8-hour observation, demonstrating the effectiveness of the alignment technique of \cite{vanweeren2026} for deep linear polarisation imaging. In the Stokes-I images with linearly polarised emission contours, the hotspots appear more prominently linearly polarised in the deeper images, and the extension denoted by `Lobe S' extends more towards the central host galaxy in the deep image. The RM value of `Hotspot S' is 6.29$\pm0.01$~rad~m$^{-2}$, while slightly lower values are measured for `Hotspot N' (6.06$\pm0.07$~rad~m$^{-2}$) and `Lobe S' (5.66$\pm0.05$~rad~m$^{-2}$). These values are consistent with what was reported by \cite{vanweeren2026}.

A more detailed study of the linear polarisation properties of the ELAIS-N1 dataset presented in this paper will be described in future work, which will include a quantification of the polarisation leakage across the field, an investigation of beam depolarisation effects, and a search for newly detected polarised sources. This builds upon the work by \cite{piras2024, piras2025} at 6\arcsec\ resolution.

\section{Discussion}\label{sec:discussion}

With the data presented in the previous sections, we compare in this section source detections across different resolutions. We conclude by touching upon a future outlook for deeper imaging with the SKA and LOFAR2.0.

\subsection{Source detections across resolutions}\label{sec:resolutioncompare}

There is no single `optimal' angular resolution to perform all types of science, as point-source sensitivity is higher at higher resolutions, whereas the surface brightness sensitivity is better at lower resolutions. Hence, some of the faint compact sources may be detected at high resolution but not at low resolution, while extended sources may fall within the detection limits of a low resolution image but remain undetected in a high resolution image. Other sources that are blended into a single source at a low resolution may be resolved into multiple components at higher resolution. 
This makes higher resolutions best suited for accurate astrometry and optical identifications (see Section \ref{sec:astrometry}), characterising and resolving detailed source structures \citep[e.g.][]{varenius2016, ramirez2018, olivencia2022, timmerman2022, mahatma2023, gloudemans2025}, or separation of different compact source components \citep[e.g. AGN vs star formation;][]{morabito2022a, morabito2025}, whereas lower resolutions are better for detecting large extended structures \citep[e.g.][]{govoni2019, vanweeren2019,vanweeren2024, botteon2020, botteon2022, dejong2022} or giant radio galaxies \citep[e.g.][]{dabhade2020, oei2024, mostert2024}.
These differences in source detections for different angular resolutions explain why the combined multi-resolution catalogue has a much higher source count compared to the individual catalogues for each image resolution (see Table \ref{table:final}).

\begin{figure}[h!]
 \centering
\includegraphics[width=0.9\linewidth]{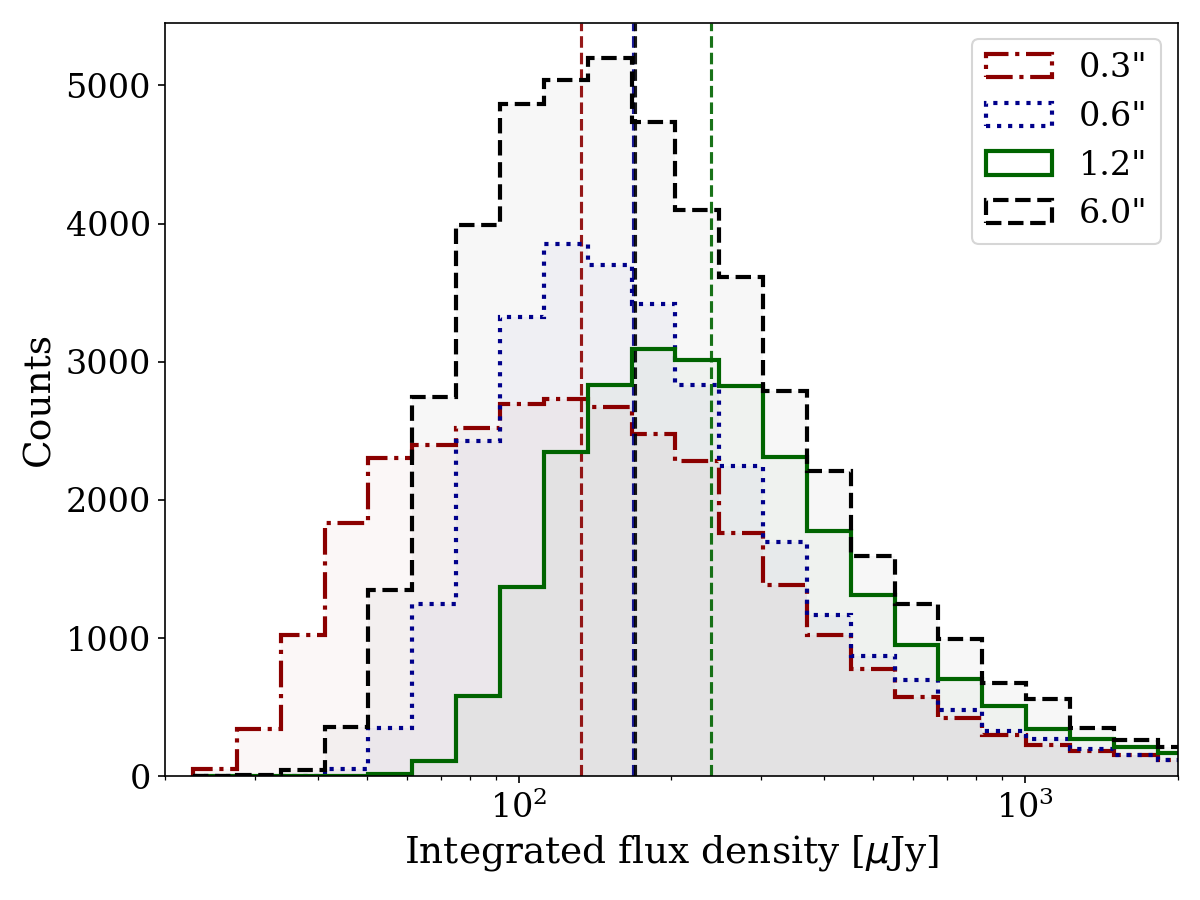}
\includegraphics[width=0.9\linewidth]{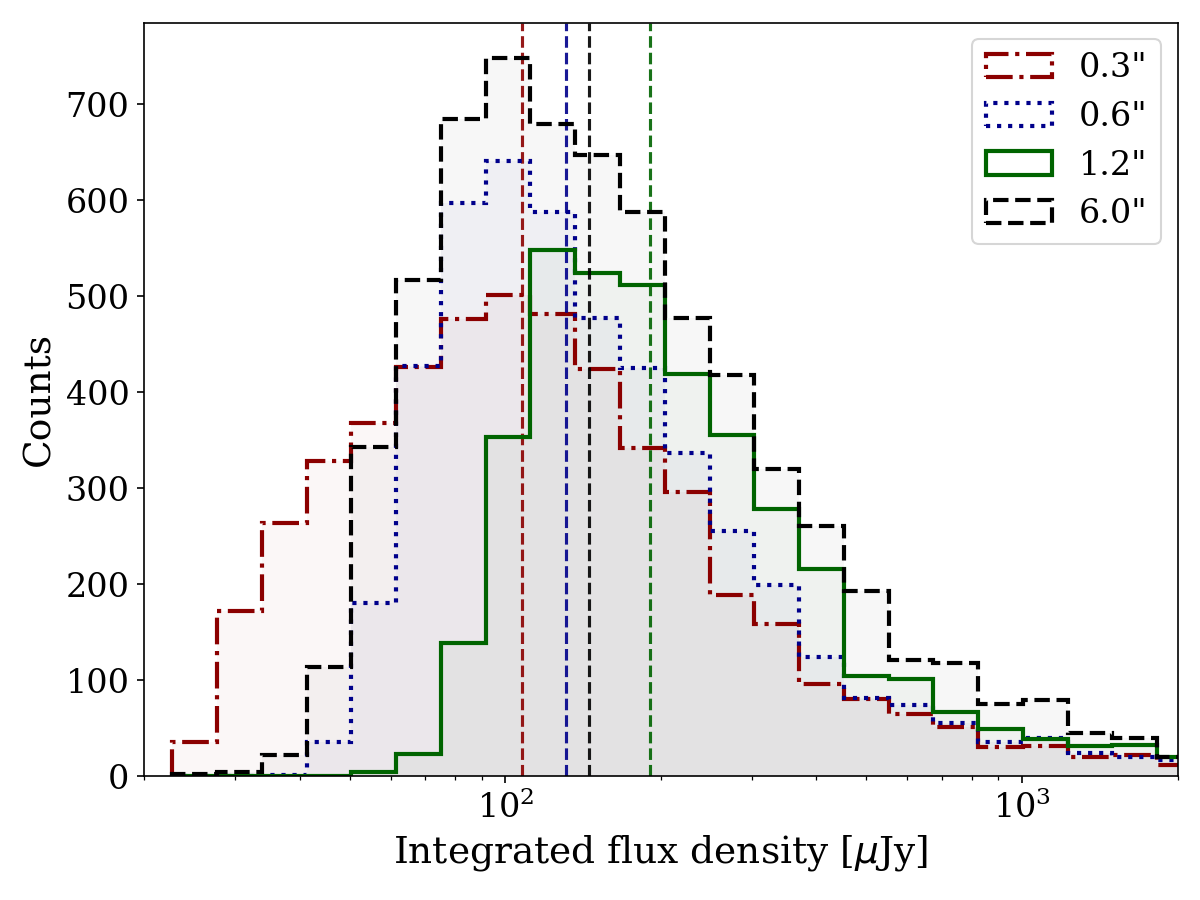}
\caption{Source count histograms as a function of integrated flux density at four angular resolutions for all unique sources (see Table~\ref{table:final}) in the full field (\textit{upper}) and within 0.5~deg of the phase centre (\textit{lower}). The 0.3\arcsec, 0.6\arcsec, and 1.2\arcsec\ resolutions correspond to the images and catalogues produced in this work, while the 6\arcsec\ sources are from the deep imaging of \citet{shimwell2025}. Dashed vertical lines mark the median integrated flux density at each resolution. Note: In the upper panel, the medians at 0.6\arcsec\ and 6\arcsec\ nearly coincide.}
\label{fig:fluxhist}
\end{figure}

Figure~\ref{fig:fluxhist} compares the apparent source counts as a function of the integrated flux density across the different angular resolutions. As expected, we find in the upper panel that the sources detected in the 0.3\arcsec\ images include the faintest radio sources, reflecting the higher point-source sensitivity at the highest resolution. However, at flux densities around 70\,$\upmu$Jy and 100\,$\upmu$Jy, the 6\arcsec\ and 0.6\arcsec\ resolution images, respectively, detect more sources than the 0.3\arcsec\ image. To limit the role of smearing in this resolution comparison, we show in the lower panel of Figure~\ref{fig:fluxhist} the same histograms for sources within 0.5~deg of the phase centre, where smearing is small (see Figure~\ref{fig:smearing}). In this region, the 6\arcsec\ and 0.6\arcsec\ images still detect more sources than the 0.3\arcsec\ image above $\sim$70\,$\upmu$Jy, while the deficit of 0.3\arcsec\ detections relative to 6\arcsec\ and 0.6\arcsec\ is only slightly smaller than for the full field. This indicates that smearing contributes but that the cross-over is also caused by the loss of surface brightness sensitivity at higher resolution, combined with the fraction of the star-forming galaxy population -- the dominant source population at these flux densities \citep{best2023} -- that is resolved at higher resolution, lowering their peak intensities below the detection threshold. For instance, at $z\sim1$, typical star-forming galaxies are large enough to be resolved at \textasciitilde0.3\arcsec\ but remain point sources at 0.6\arcsec\ \citep[e.g.][]{vanderwel2014, ward2024}, so this population may be disproportionately lost from the 0.3\arcsec\ catalogue. At the higher integrated flux density end, where radio galaxies start to become the more dominant population \citep{best2023}, the source counts are ordered by resolution, with the lower resolutions detecting more sources than higher resolutions. This is mainly due to the loss of surface brightness sensitivity towards higher resolutions, which either lowers the measured integrated flux density or causes sources to be undetected because most of their extended emission, for example from radio jets, is resolved out below our detection thresholds.

\begin{figure}[h!]
 \centering
\includegraphics[width=0.9\linewidth]{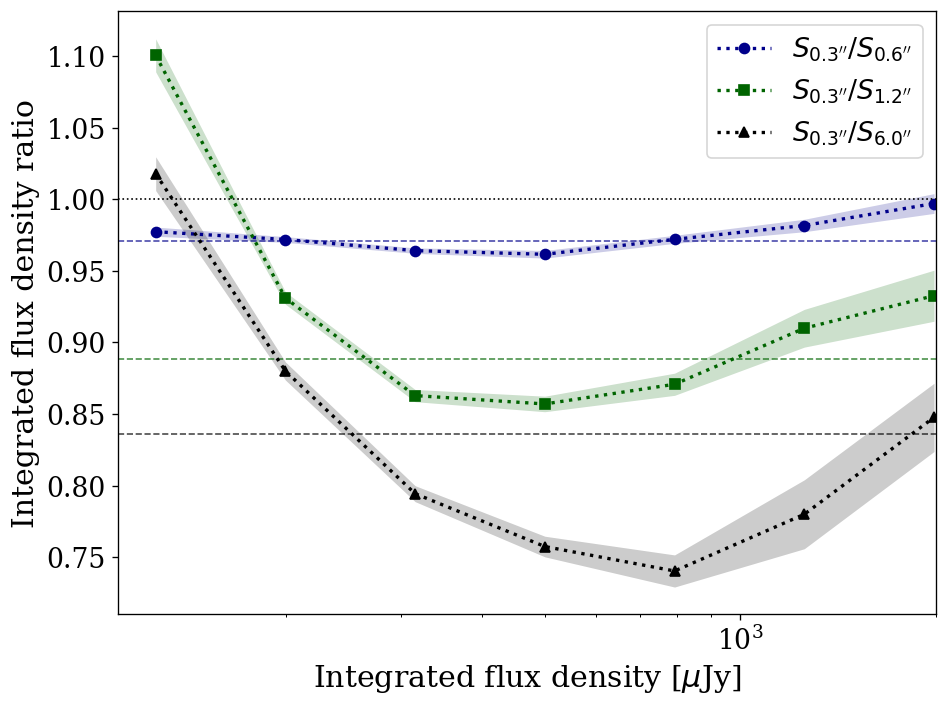}
\caption{Median ratio of the integrated flux densities at 0.3\arcsec\ and lower resolution as a function of the integrated flux density at the lower resolution for single-component sources detected at both resolutions with integrated flux densities exceeding five times the \texttt{PyBDSF} uncertainty. Shaded areas show the bootstrap uncertainties on the median in each flux density bin, and dashed horizontal lines show the median ratio over all flux densities for each pair. The dotted line marks a ratio of unity.}
\label{fig:fluxdensity_ratio}
\end{figure}

To estimate the loss of surface brightness sensitivity, we compare the median ratio of the integrated flux densities at 0.3\arcsec\ and the different lower resolutions as a function of the integrated flux density for single-component sources detected at both resolutions with integrated flux densities exceeding five times the \texttt{PyBDSF} uncertainty. The selection of single-component sources provides a lower limit of the loss of surface brightness sensitivity across the whole population of radio sources. Over all flux densities, we find median ratios of $S_{0.3}/S_{0.6}=0.97\pm0.01$, $S_{0.3}/S_{1.2}=0.89\pm0.01$, and $S_{0.3}/S_{6.0}=0.84\pm0.01$. The uncertainties are the standard error of the median, obtained by bootstrap resampling. 
The ratios decrease as the resolution difference increases, confirming the loss of surface brightness sensitivity towards higher angular resolution.

Figure~\ref{fig:fluxdensity_ratio} shows how these integrated flux density ratios vary with integrated flux density. This behaviour reflects three effects: changing source populations, the higher point-source sensitivity at higher resolutions, and the better surface brightness sensitivity at lower resolutions. At the lowest integrated flux densities, where star-forming galaxies are the dominant source population \citep{best2023}, the ratios approach or exceed unity, most strongly for the 1.2\arcsec\ comparison. This is because the 0.3\arcsec\ images have the best point-source sensitivity and the 1.2\arcsec\ images the poorest, so faint compact star-forming galaxies are better recovered at high resolution. Towards $\sim$500--800~$\upmu$Jy the ratios decline. In this regime star-forming galaxies still dominate, but the brighter ones tend to be larger, so a larger fraction of their emission may be partly resolved out in the 0.3\arcsec\ images, and so the loss of surface brightness sensitivity at high resolution has its largest effect here on the ratios. At higher flux densities the ratios rise again as radio-loud AGN become the dominant population. Because we selected single-component sources, this population is biased towards compact AGN cores, which, owing to their small angular sizes, are expected to have similar flux densities at all resolutions. The minimum of the ratio shifts to higher flux densities as the difference in resolution increases. This is partly because sources have higher integrated flux densities at lower resolution, and partly because the emission resolved out on larger angular scales comes mainly from brighter more extended sources.

\subsection{Future deeper imaging}\label{sec:future}

While confusion noise will not be a limitation for deeper imaging with LOFAR at sub-arcsecond resolutions (see Section \ref{sec:confusion}), it will become an issue for the upcoming Square Kilometre Array Low \citep[SKA-Low;][]{dewdney2009}. The SKA-Low will operate at similar frequencies to LOFAR but with significantly higher sensitivity \citep{zwart2015, prandoni2015}. This instrument is therefore expected to become confusion-limited within a few hours (depending on the array configuration), owing to its current maximum baseline length of 73.4~km.\footnote{\url{https://www.skao.int/en/explore/telescopes/SKA-Low}} As a result, SKA-Low will not be able to surpass the sensitivities achieved in this work unless its maximum baselines are extended \citep{timmerman2026}.

For LOFAR2.0, the ongoing upgrade of LOFAR \citep[e.g.][]{edler2021,orru2024,wang2025}, deeper observations are planned. One of the future planned large projects will be the LOFAR2.0 Ultra Deep Observation (LUDO; PIs: L.K. Morabito \& P.N. Best),\footnote{\url{https://sites.google.com/view/lofarultradeep/home}} which aims to image \textasciitilde3200~hours of LOFAR data of the Euclid Deep Field North (EDFN) and achieve a depth of 2~$\upmu$Jy~beam$^{-1}$ across the central 3.7~deg$^2$. This follows on the shallower LOFAR surveys of the same field at 6\arcsec~by \cite{bondi2024} and at 0.3\arcsec~by \citep{dejong2026edfn}. The combination of the deep optical and near-infrared data from the Euclid Deep Survey \citep{euclid} with ultra-deep high-resolution low-frequency radio observations from LOFAR will provide a powerful dataset for studying the relationship between radio and optical/near-infrared emission from galaxies across a wide range of redshifts \citep{bisigello2025,magliocchetti2025}.

Enabling the imaging of thousands of hours of LOFAR data within feasible wall time will benefit from further reductions in the processing cost of \texttt{PILOT}. A large part of this work will be carried out by the LOFAR Enhanced Network for Sharp Surveys (LENSS) project.\footnote{\url{https://www.astron.nl/lenss-awarded}} Potential improvements for LENSS are outlined by \cite{dejong2025b}, who highlight in particular substantial gains achievable through porting parts of the calibration and imaging pipeline to GPUs \citep[e.g.][]{tol2018, vandertol2026, veenboer2020, vdhaak2024, derubeis2025}.

\section{Summary and conclusions}\label{sec:conclusions}

We have presented the release of ultra-deep wide-field sub-arcsecond images and a multi-resolution catalogue of the ELAIS-N1 field using 200~hours of LOFAR HBA observations. Our main results are summarised as follows:

\begin{itemize}
    \item By combining recent advances in calibration, imaging, and data processing strategies, we demonstrated that imaging 400~TB of visibility data to produce high-fidelity sub-arcsecond resolution images over a $2.5\times2.5$~deg$^{2}$ field at metre wavelengths is now feasible within an acceptable computational time below 1 million CPU~core~hours.

    \item We achieved a highest central point-source sensitivity of 5.3~$\upmu$Jy~beam$^{-1}$ at a resolution of 0.33\arcsec$\times0.39$\arcsec\ and an area-weighted sensitivity of 7.4~$\upmu$Jy~beam$^{-1}$ at 0.36\arcsec$\times0.45$\arcsec, and thus the images we presented represent the deepest radio images below 1.28~GHz and are among the overall deepest radio images to date. In the centre of the field, we obtained a nearly factor-of-two improvement over the previous deepest LOFAR wide-field image of the same field but at a lower angular resolution \citep{shimwell2025} and an improvement of a factor of \textasciitilde2.5 compared to the previous data release of ELAIS-N1 at sub-arcsecond resolution \citep{dejong2024a}. After calibration, the astrometry yields an accuracy of about $\Delta$RA = $22 \pm 71$~mas\ and $\Delta$DEC = $-17 \pm 41$~mas, which by using 95 \textit{Gaia} quasars we further improved to $\Delta$RA = $-3 \pm 27$~mas\ and $\Delta$DEC = $2 \pm 17$~mas for 22 of our 24 facets. We also applied facet-to-facet flux density scale corrections to align our measurements with those of \cite{shimwell2025} at 6\arcsec~resolution.

   \item We produced a multi-resolution catalogue by combining source detections from our 0.3\arcsec, 0.6\arcsec, and 1.2\arcsec\ images with those identified in the 6\arcsec\ image of \citet{shimwell2025} within our FoV. The resulting catalogue contains 62\,916 unique sources. This is one of the largest numbers of radio source detections from a single telescope pointing, highlighting LOFAR's capability for ultra-deep imaging across large FoVs. The combination of high source density and sub-arcsecond resolution makes this catalogue unique compared to any other radio catalogue.

    \item We showed that time and bandwidth smearing remain a limitation at the highest angular resolution, particularly towards the edges of the FoV, where peak intensities are reduced by up to 80\% of their intrinsic values at 0.3\arcsec. This has a disproportionate effect on the completeness of source detections at 0.3\arcsec~compared to the other lower resolutions. A significant contribution to this effect arises from additional time averaging in roughly half of the ELAIS-N1 observations. The smearing effects introduced by ionospheric conditions and by SVA to combine observations in visibility space are comparatively small.

    \item In contrast to lower-resolution imaging with LOFAR's Dutch-only baselines and future SKA-Low observations, our high-resolution image is not limited by confusion noise and will not become confusion limited with increasing observing time. This provides a unique capability to explore increasingly faint radio source populations while having the resolution to resolve jets and lobes and separate compact sources and components.

    \item After leakage calibration, we measured linear polarisation leakages of 0.12\% and circular polarisation leakages of 0.02\% on the leakage calibrator, which is about a factor of two better than the typical leakage in LoTSS at 6\arcsec\ resolution. We also recovered more linear polarisation across the lobe from the polarisation angle alignment calibrator. This demonstrates the potential for future deep polarisation studies at sub-arcsecond resolution, enabling the detection of fainter polarised emission with increasing total integration time.

\end{itemize}
Our results establish a new benchmark for low-frequency radio imaging in terms of the combination of depth, resolution, and survey area. In addition, this work lays the foundation for future ultra-deep surveys with LOFAR2.0, such as the planned LUDO survey, which will extend these techniques to the imaging of several thousand hours of LOFAR observations. Overall, this work demonstrates that ultra-deep high-resolution imaging at metre wavelengths with LOFAR is now reaching a mature and scalable regime and thus opening new capabilities for studying the deep radio sky at the lowest frequencies.

\section*{Data availability}

The input visibility data that were processed in this work are publicly available from the LOFAR LTA (\url{https://lta.lofar.eu}). The images and catalogues produced in this work have been deposited, with data provenance, in the SURF Data Repository (SDR) and can be accessed without restrictions or registration at \url{https://repository.surf.nl/records/xqxgr-cdf21}. The SDR guarantees a retention period of at least 5 years, after which the retention period will be extended or the data will be moved to another public repository. The catalogues and images will also be made available through the ASTRON Virtual Observatory service, \url{https://vo.astron.nl}. The data have been enriched with metadata to improve reproducibility. These efforts are part of the OSCARS project \citep{dejong_fair} and ensure that the resulting data products comply with the FAIR principles \citep{wilkinson2016, oscars}.

\begin{acknowledgement}

The authors thank C. Schreiber for useful conversations about cataloguing. The authors acknowledge the OSCARS project, which has received funding from the European Commission’s Horizon Europe Research and Innovation programme under grant agreement No. 101129751. RT is grateful for support from the UKRI Future Leaders Fellowship (grant MR/Y020405/1). This work was supported by the STFC [grants ST/T000244/1, ST/V002406/1].
 FS appreciates the support of STFC [ST/Y004159/1].
RK acknowledges support from the Leverhulme Trust through a Leverhulme Early Career Fellowship. 
CLH acknowledges support from the Science and Technology Facilities Council (STFC) through grant ST/Y000951/1.

This work is part of the project CORTEX (NWA.1160.18.316) of the research programme NWA-ORC which is (partly) financed by the Dutch Research Council (NWO). This work is co-funded by the EGI-ACE project (Horizon 2020) under Grant number 101017567.
This publication is part of the project \textit{Deep high-resolution LOFAR imaging} with file number 2023.040 of the research programme Computing Time on National Computing Facilities which is (partly) financed by the Dutch Research Council (NWO).
This publication is part of the project \textit{Uncovering the wide high-resolution sky with LOFAR} with file number \textit{2025.043} of the research programme \textit{Computing Time on National Computer Facilities} which is (partly) financed by the Dutch Research Council (NWO) under the grant \url{ https://doi.org/10.61686/JACQA20438}.
This publication is part of the project Probing the Formation of Galaxy Clusters and their Progenitors with High-Resolution Radio Imaging with file number VI.C.252.058 of the NWO Talent Programme which is (partly) financed by the Dutch Research Council (NWO) under the grant \url{https://doi.org/10.61686/AZVOB10686}.

The public release of the data associated with this work was made possible by funding from the Astrophysics Centre for Multi-messenger studies (ACME) project. ACME project has received funding from the European Union's Horizon Europe Research and Innovation programme under Grant Agreement No 101131928.

LOFAR is the Low Frequency Array designed and constructed by ASTRON \citep{haarlem2013}. It has observing, data processing, and data storage facilities in several countries, which are owned by various parties (each with their own funding sources), and which are collectively operated by the LOFAR ERIC under a joint scientific policy. The LOFAR resources have benefited from the following recent major funding sources: CNRS-INSU, Observatoire de Paris and Université d'Orléans, France; BMBF, MIWF-NRW, MPG, Germany; Science Foundation Ireland (SFI), Department of Business, Enterprise and Innovation (DBEI), Ireland; NWO, The Netherlands; The Science and Technology Facilities Council, UK; Ministry of Science and Higher Education, Poland; The Istituto Nazionale di Astrofisica (INAF), Italy.

This research made use of the Dutch national e-infrastructure with support of the SURF Cooperative (e-infra 180169) and the LOFAR e-infra group. The J\"ulich LOFAR Long Term Archive and the German LOFAR network are both coordinated and operated by the Jülich Supercomputing Centre (JSC), and computing resources on the supercomputer JUWELS at JSC were provided by the Gauss Centre for Supercomputing e.V. (grant CHTB00) through the John von Neumann Institute for Computing (NIC).

This research made use of the University of Hertfordshire high-performance computing facility and the LOFAR-UK computing facility located at the University of Hertfordshire and supported by STFC [ST/P000096/1], and of the Italian LOFAR IT computing infrastructure supported and operated by INAF, and by the Physics Department of Turin University (under an agreement with Consorzio Interuniversitario per la Fisica Spaziale) at the C3S Supercomputing Centre, Italy.

This research is part of the project LOFAR Data Valorization (LDV) [project numbers 2020.031, 2022.033, and 2024.047] of the research programme Computing Time on National Computer Facilities using SPIDER that is (co-)funded by the Dutch Research Council (NWO), hosted by SURF through the call for proposals of Computing Time on National Computer Facilities. 

This work made use of Flocs (\url{https://github.com/FLOCSoft}) for (part of) its data reduction.

\end{acknowledgement}

\bibliographystyle{bst}
\bibliography{bib}

\appendix

\section{Observations}\label{app:observations}

Table \ref{table:metadata} lists the LOFAR observations that were selected for this work. 
All calibrator IDs correspond to 10-minute observations of calibrator 3C~295. The number of international stations is given in brackets. The reported numbers of stations and frequency channels correspond to the data after flagging. The RMS noise values are measured from the 1.2\arcsec~wide-field images and reflect the combination of flagging and ionospheric conditions.

\begin{table*}[htbp]
\caption{Metadata from the 25 ELAIS-N1 observations used in this paper.}
\centering
\begin{tabular}{l c c c c c c} \toprule
    \textbf{Observ. ID} & \textbf{Project} & \textbf{Calibrator ID} & \textbf{Observ. date} & \textbf{Stations (int)} & \textbf{Frequencies} & 
    \textbf{RMS 1.2\arcsec} \\ 
    & & & & & (MHz) & ($\upmu$Jy beam$^{-1}$) \\ \midrule
    
    L686892 & LT10\_012 & L686888 & 06-12-2018 & 70 (12) & 120-165 & 116 \\
    
    L686906 & LT10\_012 & L686902 & 08-12-2018 & 75 (13) & 120-165 & 106 \\
    
    L686920 & LT10\_012 & L686916 & 03-12-2018 & 74 (12) & 120-165 & 102 \\
    
    L686934 & LT10\_012 & L686940 & 07-12-2018 & 74 (12) & 120-165 & 110 \\
    
    L686948 & LT10\_012 & L686944 & 05-12-2018 & 73 (11) & 120-167 & 113 \\
    
    L686962 & LT10\_012 & L686958 & 26-11-2018 & 73 (11) & 120-165 & 88 \\
    
    L686976 & LT10\_012 & L686972 & 25-11-2018 & 73 (11) & 128-167 & 112 \\
    
    L686990 & LT10\_012 & L686986 & 24-11-2018 & 73 (11) & 126-167 & 103 \\

    L687004 & LT10\_012 & L687000 & 22-11-2018 & 70 (9) & 128-167 & 106 \\
    
    L760853 & LT10\_012 & L760849 & 04-12-2019 & 67 (12) & 120-166 & 100 \\
    
    L769393 & LT10\_012 & L769389 & 25-05-2020 & 75 (13) & 120-165 & 89 \\
    
    L769407 & LT10\_012 & L769403 & 19-05-2020 & 71 (12) & 120-165 & 84 \\
    
    L769421 & LT10\_012 & L769417 & 12-05-2020 & 73 (13) & 120-167 & 126 \\
    
    L769435 & LT10\_012 & L769431 & 15-03-2020 & 72 (11) & 120-166 & 96\\

    L798022 & LT14\_003 & L798030 & 12-11-2020 & 72 (11) & 117-166 & 92 \\
    
    L798074 & LT14\_003 & L798082 & 14-11-2020 & 74 (11) & 115-164 & 90 \\

    L798092 & LT14\_003 & L798100 & 20-11-2020 & 68 (11) & 115-164 & 95 \\

    L798128 & LT14\_003 & L798136 & 26-11-2020 & 69 (9) & 115-166 & 81 \\

    L811718 & LT14\_003 & L811726 & 19-03-2021 & 72 (11) & 117-165 & 84 \\

    L811736 & LT14\_003 & L811744 & 20-03-2021 & 75 (13) & 115-164 & 92 \\

    L816272 & LT14\_003 & L816280 & 15-05-2021 & 75 (14) & 115-164 & 93 \\

    L816308 & LT14\_003 & L816316 & 23-05-2021 & 75 (13) & 116-162 & 86 \\

    L816344 & LT14\_003 & L816352 & 27-05-2021 & 73 (14) & 115-166 & 92 \\

    L818954 & LT14\_003 & L818962 & 22-06-2021 & 72 (10) & 115-156 & 117 \\
    
    L833466 & LT14\_003 & L833474 & 09-10-2021 & 73 (13) & 119-165 & 115 \\
   \bottomrule
\end{tabular}
\label{table:metadata}
\end{table*}

\section{Imaging properties of individual facets}\label{app:facetproperties}

We provide in Table~\ref{table:facetinformation} the restoring beam sizes and the RMS noise levels measured for each of the 24 individual facets at the three imaging resolutions considered in this work. The values were obtained from the corresponding primary-beam-corrected residual images, with the RMS estimated using iterative $3\sigma$ clipping about the median. The properties of the final mosaics are also included for comparison.

\begin{table*}
\caption{Restoring beam resolution, as fitted by \texttt{WSClean}, and RMS noises of each individual facet and for the mosaic.}
\centering
\begin{tabular}{|c| c c | c c | c c |}
\arrayrulecolor{gray}\toprule
 & \multicolumn{2}{c}{{\large\textbf{0.3\arcsec}}} 
 & \multicolumn{2}{c}{{\large\textbf{0.6\arcsec}}} 
 & \multicolumn{2}{c}{{\large\textbf{1.2\arcsec}}} \\ \midrule
{\small \textcolor{darkgray}{\textbf{Facet}}} 
& {\small \textcolor{darkgray}{\textbf{Resolution}}} 
& {\small \textcolor{darkgray}{\textbf{RMS}}} 
& {\small \textcolor{darkgray}{\textbf{Resolution}}} 
& {\small \textcolor{darkgray}{\textbf{RMS}}} 
& {\small \textcolor{darkgray}{\textbf{Resolution}}} 
& {\small \textcolor{darkgray}{\textbf{RMS}}} \\
& {\small (arcsec$^{2}$)} 
& {\small ($\upmu$Jy~beam$^{-1}$)} 
& {\small (arcsec$^{2}$)} 
& {\small ($\upmu$Jy~beam$^{-1}$)} 
& {\small (arcsec$^{2}$)} 
& {\small ($\upmu$Jy~beam$^{-1}$)} \\ [0.15cm] \midrule
\textbf{0}  & 0.35$\times$0.41 & 7.2  & 0.56$\times$0.60 & 9.9  & 1.08$\times$2.12 & 15.5 \\
\textbf{1}  & 0.35$\times$0.42 & 8.1  & 0.57$\times$0.61 & 11.1 & 1.12$\times$2.38 & 18.2 \\
\textbf{2}  & 0.33$\times$0.39 & 5.3  & 0.54$\times$0.57 & 7.7  & 1.00$\times$1.50 & 13.2 \\
\textbf{3}  & 0.33$\times$0.39 & 5.4  & 0.54$\times$0.57 & 8.1  & 1.00$\times$1.48 & 14.7 \\
\textbf{4}  & 0.33$\times$0.39 & 6.2  & 0.55$\times$0.58 & 8.9  & 1.02$\times$1.66 & 15.0 \\
\textbf{5}  & 0.35$\times$0.42 & 7.6  & 0.57$\times$0.60 & 10.3 & 1.09$\times$2.24 & 16.7 \\
\textbf{6}  & 0.33$\times$0.40 & 5.6  & 0.55$\times$0.58 & 8.3  & 1.02$\times$1.66 & 14.0 \\
\textbf{7}  & 0.35$\times$0.42 & 8.6  & 0.57$\times$0.61 & 11.1 & 1.10$\times$2.27 & 17.5 \\
\textbf{8}  & 0.37$\times$0.44 & 8.8  & 0.62$\times$0.63 & 12.3 & 1.29$\times$2.76 & 18.6 \\
\textbf{9}  & 0.34$\times$0.41 & 6.5  & 0.55$\times$0.59 & 9.2  & 1.04$\times$1.82 & 15.0 \\
\textbf{10} & 0.36$\times$0.46 & 10.3 & 0.61$\times$0.65 & 13.9 & 1.37$\times$2.91 & 20.9 \\
\textbf{11} & 0.36$\times$0.46 & 10.6 & 0.62$\times$0.67 & 14.3 & 1.44$\times$2.94 & 19.3 \\
\textbf{12} & 0.36$\times$0.43 & 7.8  & 0.59$\times$0.61 & 11.8 & 1.11$\times$2.38 & 16.3 \\
\textbf{13} & 0.35$\times$0.42 & 7.4  & 0.58$\times$0.61 & 10.1 & 1.10$\times$2.32 & 16.5 \\
\textbf{14} & 0.35$\times$0.43 & 8.3  & 0.58$\times$0.62 & 11.3 & 1.14$\times$2.46 & 17.8 \\
\textbf{15} & 0.35$\times$0.42 & 7.6  & 0.57$\times$0.61 & 10.8 & 1.07$\times$2.19 & 17.5 \\
\textbf{16} & 0.38$\times$0.45 & 10.0 & 0.64$\times$0.65 & 13.3 & 1.20$\times$2.68 & 18.6 \\
\textbf{17} & 0.34$\times$0.41 & 6.9  & 0.56$\times$0.59 & 8.8  & 1.03$\times$1.86 & 14.4 \\
\textbf{18} & 0.34$\times$0.40 & 5.4  & 0.55$\times$0.58 & 8.1  & 0.99$\times$1.50 & 13.3 \\
\textbf{19} & 0.35$\times$0.43 & 8.1  & 0.58$\times$0.61 & 11.2 & 1.13$\times$2.46 & 17.6 \\
\textbf{20} & 0.34$\times$0.40 & 6.6  & 0.55$\times$0.59 & 9.4  & 1.04$\times$1.83 & 15.4 \\
\textbf{21} & 0.36$\times$0.44 & 8.2  & 0.59$\times$0.63 & 12.1 & 1.24$\times$2.69 & 19.3 \\
\textbf{22} & 0.34$\times$0.41 & 7.3  & 0.56$\times$0.59 & 10.4 & 1.05$\times$1.99 & 16.2 \\
\textbf{23} & 0.33$\times$0.39 & 5.4  & 0.55$\times$0.58 & 8.4  & 1.01$\times$1.62 & 14.2 \\
\midrule
\textbf{Mosaic} & $0.36\times0.45$ & 7.4 & $0.58\times0.62$ & 10.3 & $1.0\times1.5$ & 16.5 \\
\bottomrule
\end{tabular}
\tablefoot{Facet numbers correspond to the numbers shown in Figure~\ref{fig:facet_layout}.}
\label{table:facetinformation}
\end{table*}

\section{Dutch versus European LOFAR}\label{appendix:6vs0.3}

\begin{figure*}[htbp]
 \centering
 \includegraphics[width=0.7\linewidth]{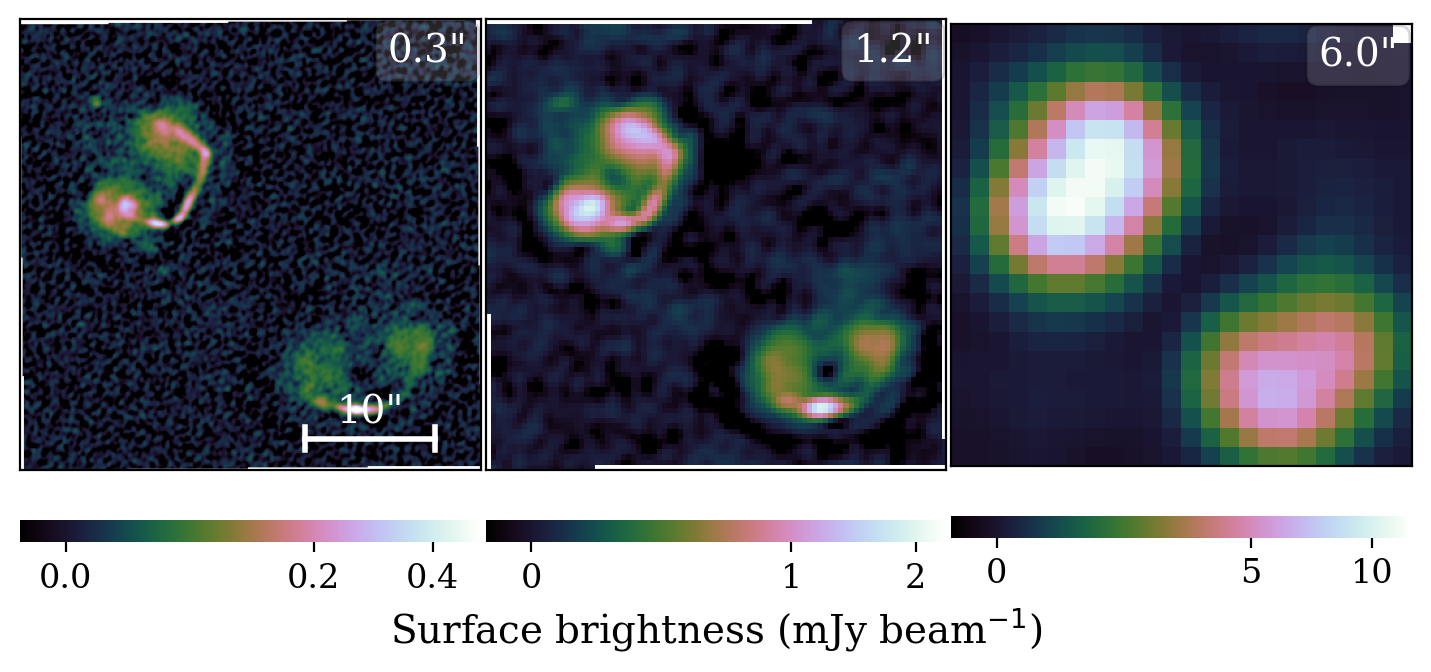}
 \includegraphics[width=0.7\linewidth]{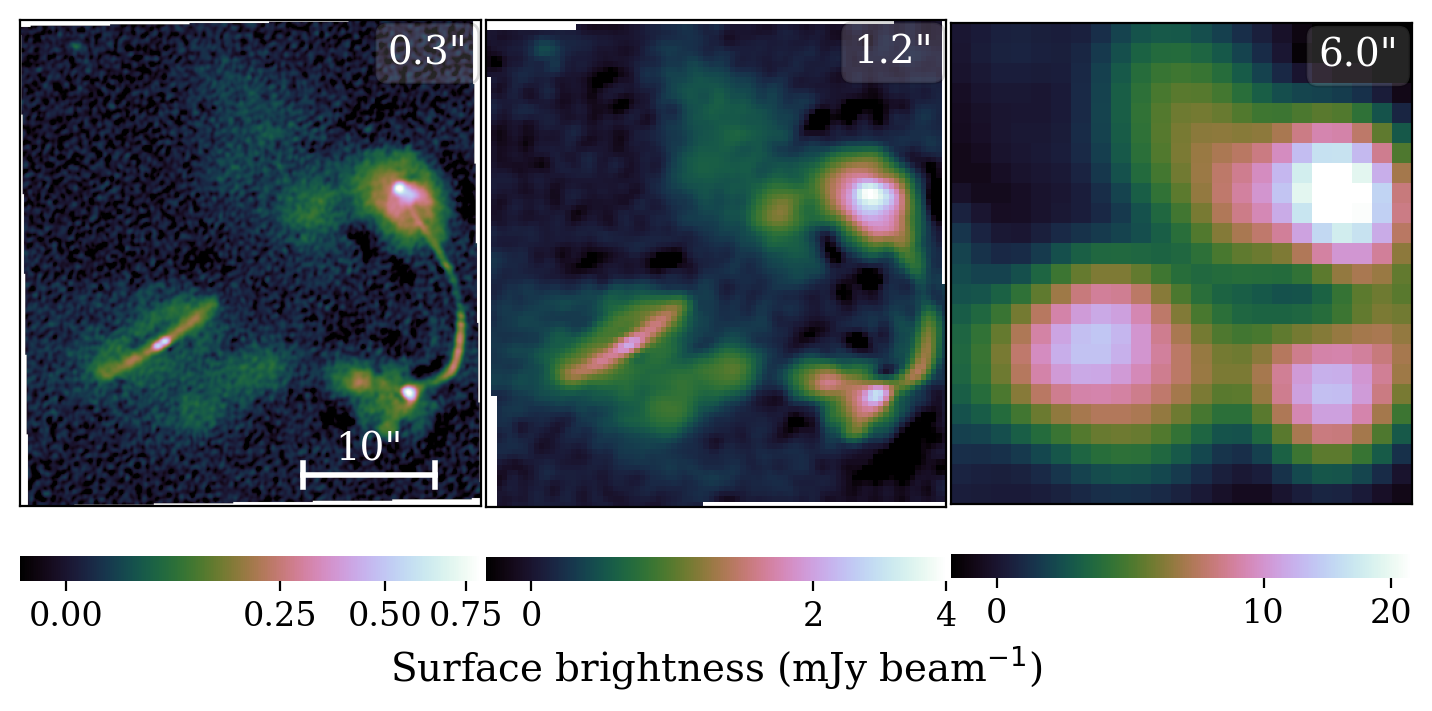}
 \includegraphics[width=0.7\linewidth]{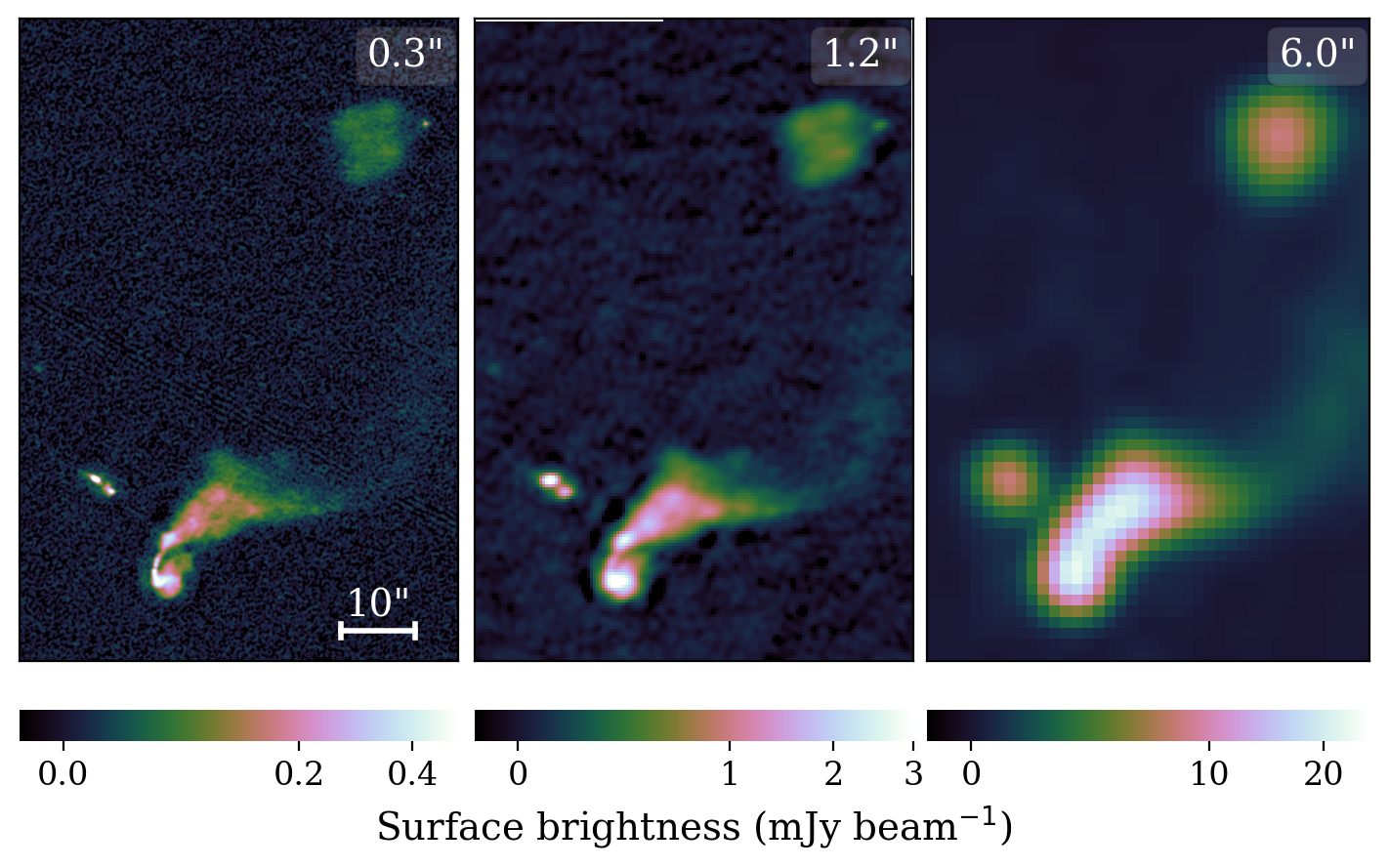}
  \caption{Comparison between 0.3\arcsec, 1.2\arcsec, and 6\arcsec~resolution images for different sky areas containing various resolved radio galaxies.}
\label{fig:6vs0.3}
\end{figure*}

To demonstrate the increased resolving power of the full European LOFAR at different resolutions, Figure~\ref{fig:6vs0.3} presents selected regions imaged for the 0.3\arcsec, 1.2\arcsec, and 6\arcsec~images. These comparisons highlight how radio galaxies that appear as compact, blob-like sources at 6\arcsec~are fully resolved into complex morphologies at sub-arcsecond resolution. With the depth achieved from 200~hours of LOFAR observations, most of the faint extended emission seen at 6\arcsec~and 1.2\arcsec~resolution is also recovered in the 0.3\arcsec\ images. This demonstrates that much of the low surface brightness emission is at these depths also captured at high resolutions.

\section{Flux density scale across wavelengths}\label{appendix:spectra}

\begin{figure*}[htbp]
 \centering
\includegraphics[width=0.33\linewidth]{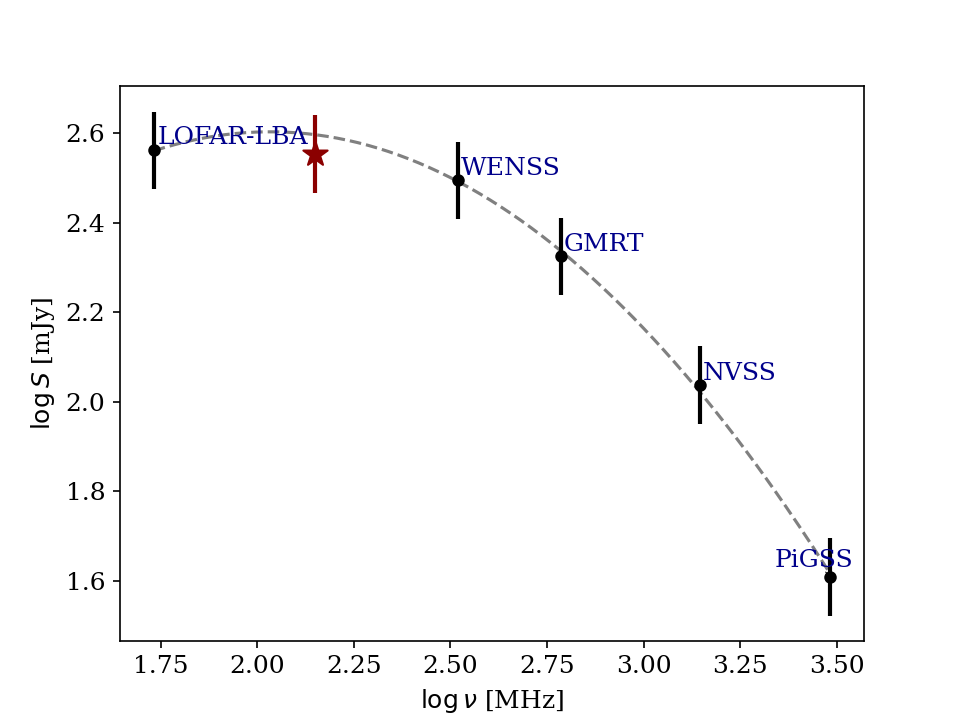}
\includegraphics[width=0.33\linewidth]{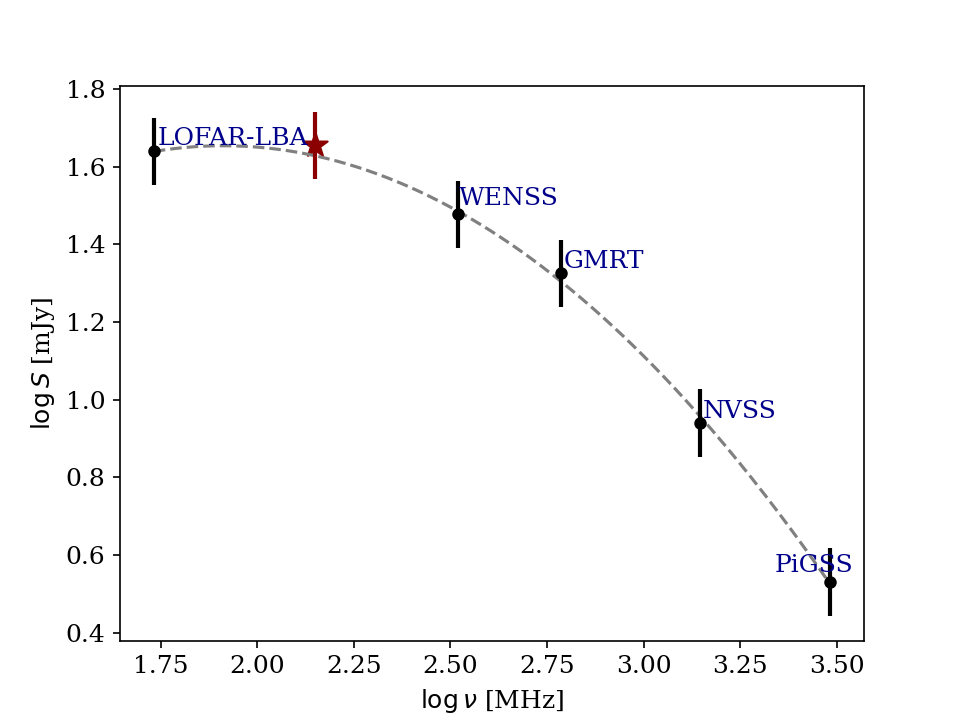}
\includegraphics[width=0.33\linewidth]{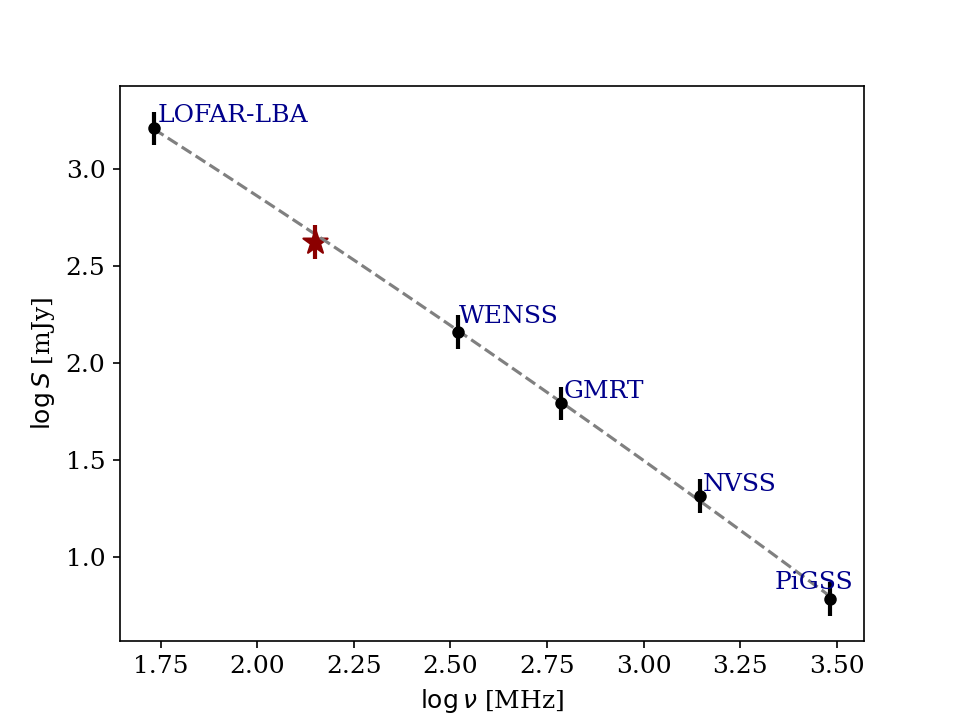}
\includegraphics[width=0.33\linewidth]{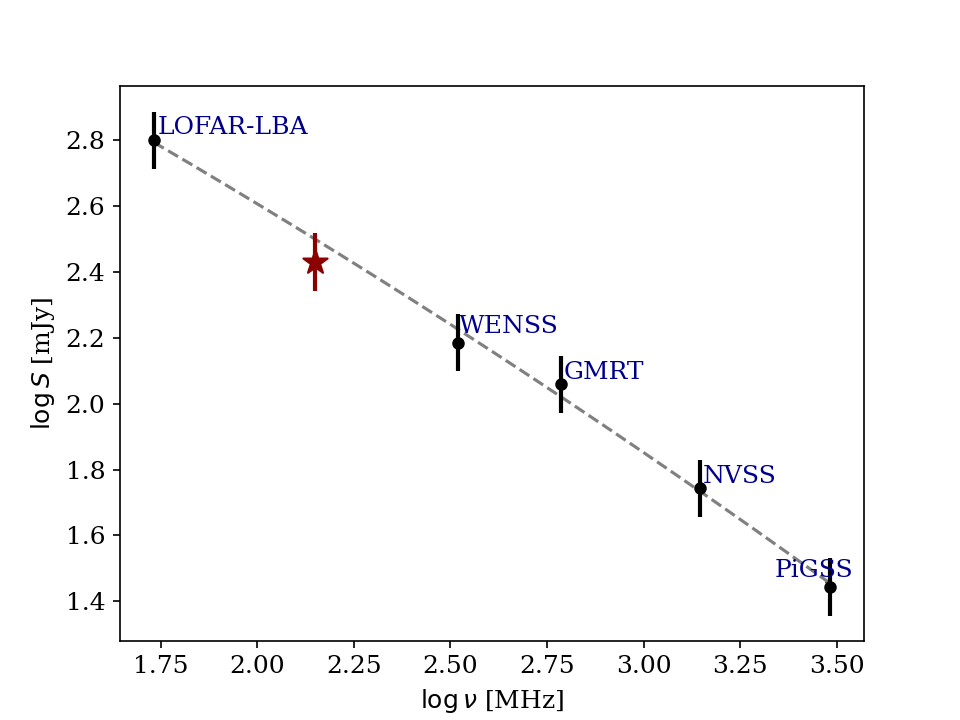}
\includegraphics[width=0.33\linewidth]{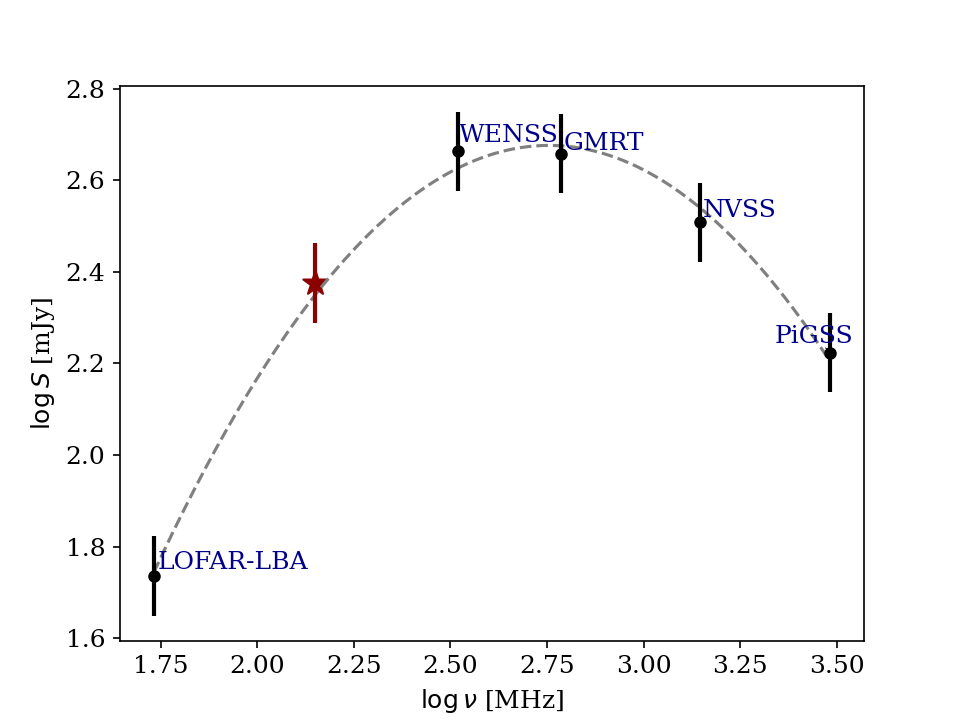}
\includegraphics[width=0.33\linewidth]{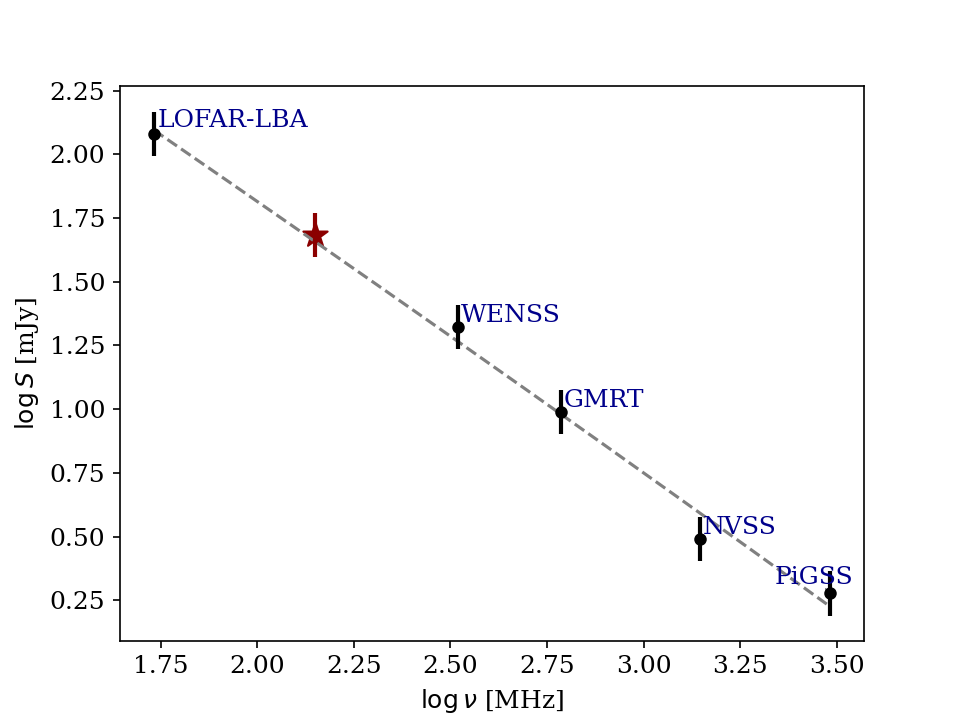}
\includegraphics[width=0.33\linewidth]{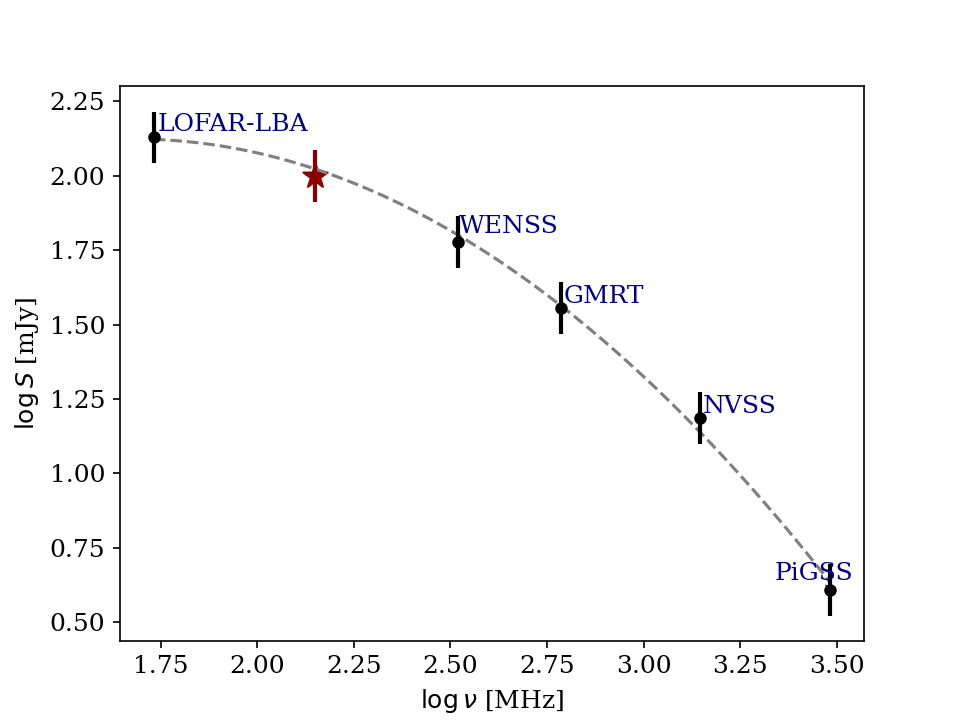}
\includegraphics[width=0.33\linewidth]{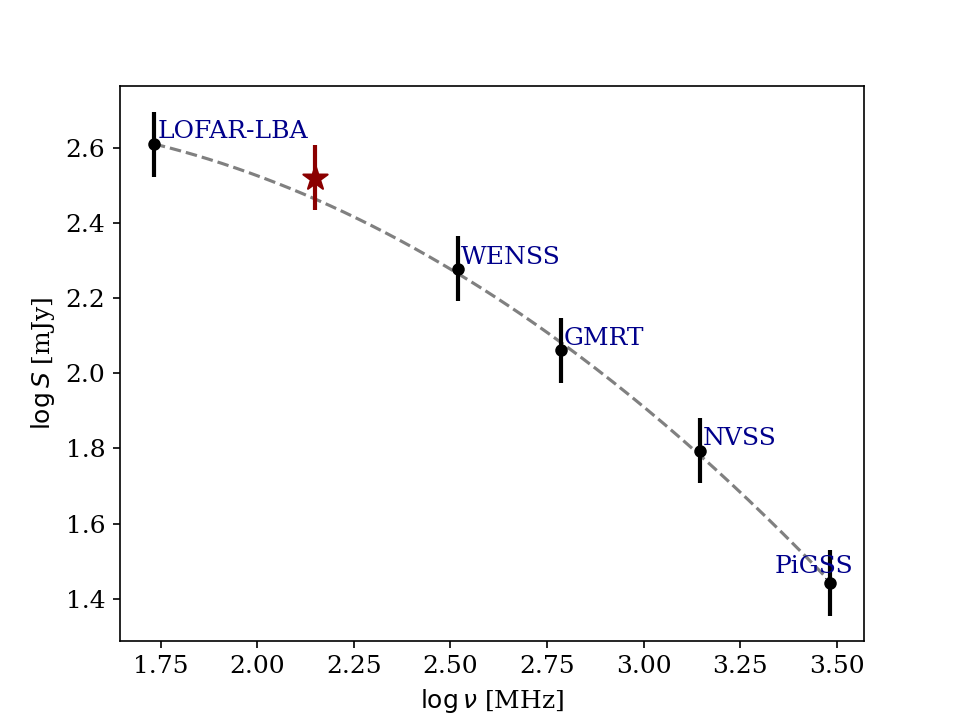}
\includegraphics[width=0.33\linewidth]{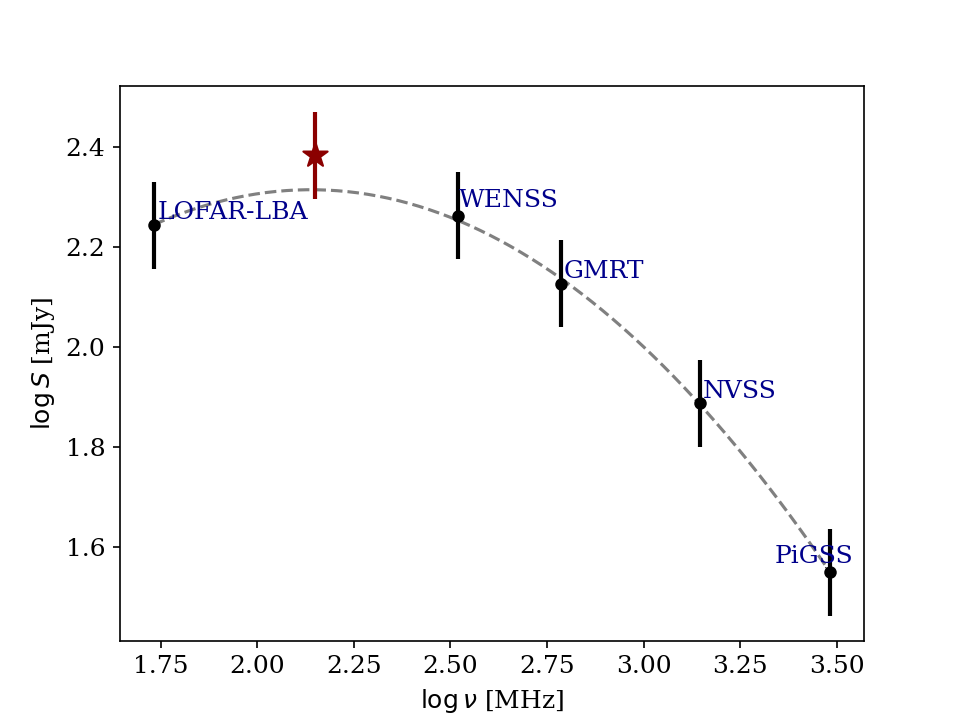}
\includegraphics[width=0.33\linewidth]{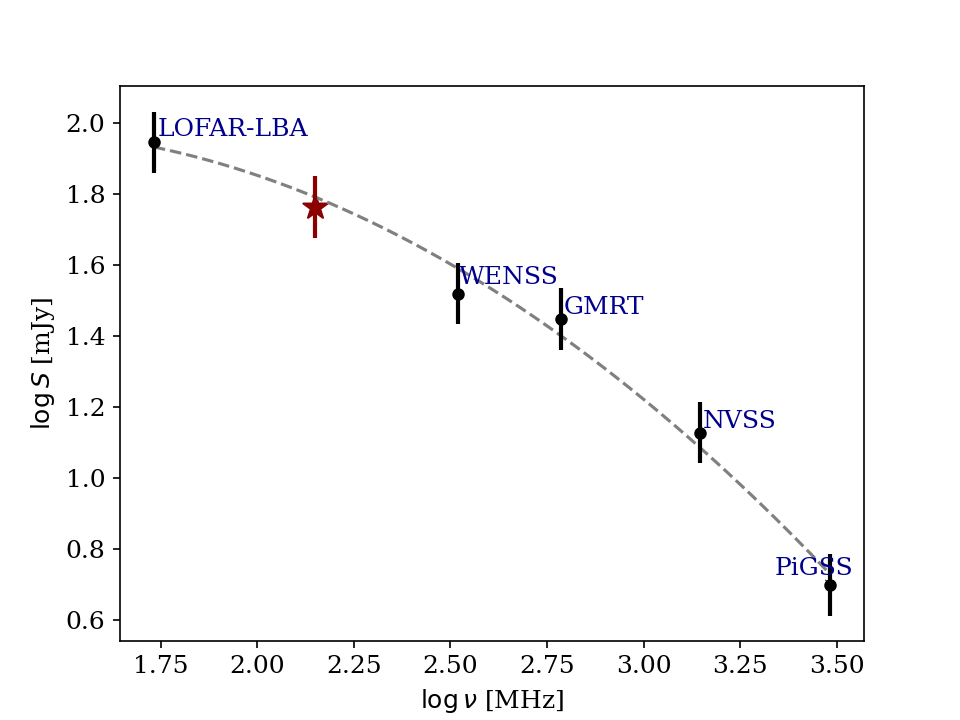}
\includegraphics[width=0.33\linewidth]{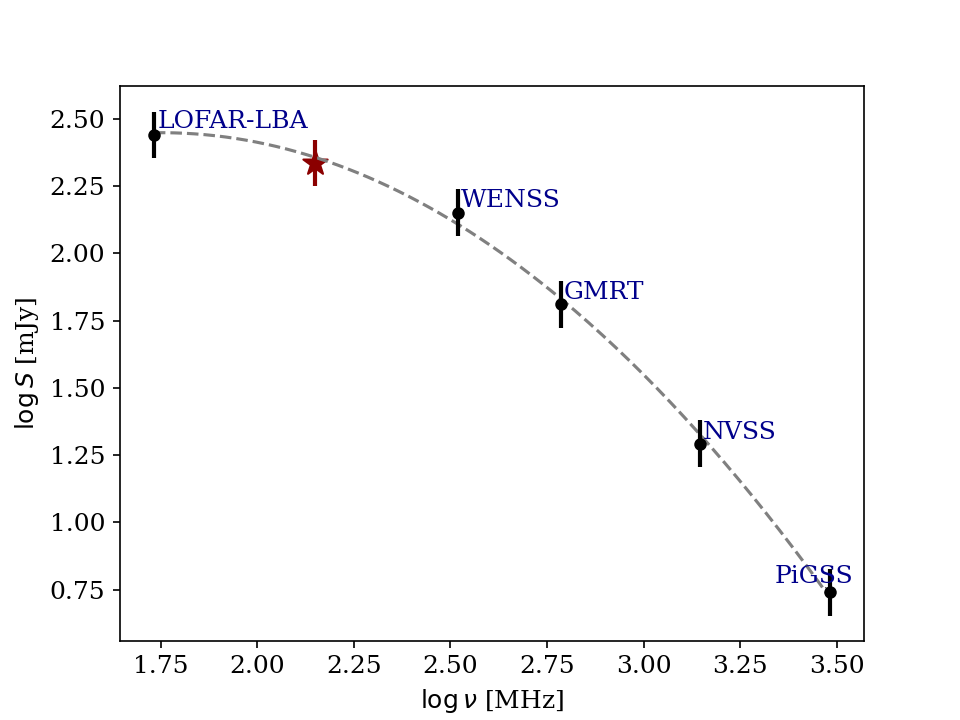}
\includegraphics[width=0.33\linewidth]{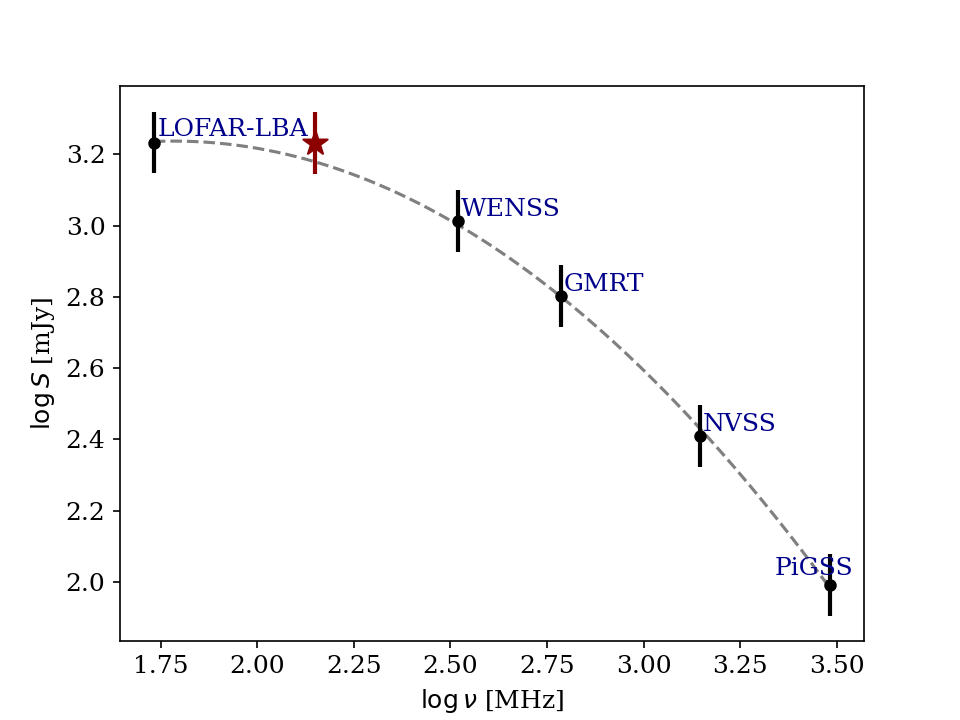}
  \caption{Fitted spectra of bright compact sources across the field after flux density scaling (see Section \ref{sec:fluxdensityscaling}). The red star corresponds to the ELAIS-N1 data presented in this work.}
\label{fig:spectra1}
\end{figure*}

In Figure \ref{fig:spectra1} we plot the spectra of several compact sources across the ELAIS-N1 FoV to demonstrate the accuracy of our final flux density scale. We used data from the Pi GHz Sky Survey \citep[PiGSS][]{croft2013}, NRAO VLA Sky Survey at 1.4~GHz \citep[NVSS;][]{condon1998}, the GMRT at 610~MHz \citep{garn2008}, the Westerbork Northern Sky Survey at 325~MHz \citep[WENSS;][]{rengelink1997}, and an image obtained after reducing data from an LBA observation at 54~MHz from the ELAIS-N1 field with the \texttt{LiLF}\footnote{\url{https://github.com/revoltek/LiLF}} calibration pipeline \citep{degasperin2018, degasperin2019, degasperin2020}. We also adopt a conservative uncertainty of 20\% on the integrated flux density scale, following LoTSS \citep{shimwell2022}.

\section{Smearing reductions}\label{appendix:smearing}

Time and bandwidth smearing remain significant issues (see Section~\ref{sec:smearing}), particularly for science cases where the recovery of accurate morphologies is essential, such as weak lensing \citep[e.g.][]{liu2025} or precise source size measurements \citep[e.g.][]{sweijen2025}. One straightforward strategy for reducing smearing would be to increase the time and frequency resolution of the data in the LTA by a factor of two. This would improve the maximum peak-intensity recovery at 0.3\arcsec\ from 50\% to 85\% \citep{bridle1999}. However, this change would also increase the number of visibilities by a factor of four. Therefore, adopting baseline-dependent averaging (BDA; e.g. \citealt{cotton1986, cotton2009, wijnholds2018}), in which shorter baselines are averaged more than longer ones, throughout the data-processing pipeline would be in this case essential \citep[e.g.][]{atemkeng2018, dejong2025b}. Enabling BDA in combination with Dynamical Statistical Compression \citep[Dysco;][]{offringa2016} or Simulated Signal Compression \citep[Sisco;][]{offringa2026} is also important to keep data volumes manageable. 

To reduce the remaining smearing, the imaging software could be extended to apply smearing corrections by modelling the smeared PSF. Such corrections are already implemented in several existing imagers \citep{noordam2010, tasse2018}. Applying such corrections on a per-facet basis could mitigate the smearing at the boundaries of each facet, observed in the left panel of Figure~\ref{fig:smearing}. More comprehensive PSF modelling that accounts for both the intrinsic smearing present in the input data and the additional averaging introduced during the construction of the facet datasets could further reduce smearing across the full image.

\end{document}